\RequirePackage{fix-cm}
\documentclass[smallextended]{svjour3}       % onecolumn (second format)
\smartqed  % flush right qed marks, e.g. at end of proof
\usepackage{graphicx}
\usepackage{algorithmic}
\usepackage{algorithm}
\usepackage{subfigure}
\usepackage{natbib}
\usepackage{url}
\usepackage{multirow}
\begin{document}

\title{The Academic Social Network
%\thanks{Our corresponding web service mentioned in this article can be accessed at http://pubstat.org}
}

\author{Tom Z. J. Fu \and Qianqian Song \and Dah Ming Chiu}

\authorrunning{Fu et al} % if too long for running head

\institute{Tom Z. J. Fu\at
              Illinois at Singapore Pte Ltd,
              Advanced Digital Sciences Center (ADSC)\\
              1 Fusionopolis Way, \#08-10 Connexis North Tower,
              Singapore 138632\\
              Tel.: +65-65919093\\
              Fax: +65-65919091\\
              \email{fuzhengjia@gmail.com}           %  \\
           \and
              Qianqian Song\at
              \email{songqianqian713@gmail.com}             %\\
           \and
              Dah Ming Chiu\at
              Room 836, Ho Sin Hang Engineering Building,\\
              Department of Information Engineering \\
              The Chinese University of Hong Kong, Shatin, N.T. Hong Kong\\
              Tel.: +852-39438357\\
              Fax: +852-26035032\\
              \email{dmchiu@ie.cuhk.edu.hk}
}

\date{Received: date / Accepted: date}

\maketitle

\begin{abstract}
Through academic publications, the authors of these publications
form a social network. Instead of sharing casual thoughts and photos
(as in Facebook), authors pick co-authors and reference papers
written by other authors. Thanks to various efforts (such as
Microsoft Libra and DBLP), the data necessary for analyzing the
academic social network is becoming more available on the Internet.
What type of information and queries would be useful for users to
find out, beyond the search queries already available from services
such as Google Scholar? In this paper, we explore this question by
defining a variety of ranking metrics on different entities -
authors, publication venues and institutions. We go beyond
traditional metrics such as paper counts, citations and h-index.
Specifically, we define metrics such as \emph{influence},
\emph{connections} and \emph{exposure} for authors. An author gains
influence by receiving more citations, but also citations from
influential authors. An author increases his/her connections by
co-authoring with other authors, and specially from other authors
with high connections. An author receives exposure by publishing in
selective venues where publications received high citations in the
past, and the selectivity of these venues also depends on the
influence of the authors who publish there. We discuss the
computation aspects of these metrics, and similarity between
different metrics.
%% newly added start
With additional information of author-institution relationships, we
are able to study institution rankings based on the corresponding
authors' rankings for each type of metric as well as different
domains.
%% newly added end
We are prepared to demonstrate these ideas with a web site
(http://pubstat.org) built from millions of publications and
authors.

%\keywords{Data Cleaning \and Academic Social Network \and Paper Citation}
%%
%%%%%%%%%%%%%%%%%%%%%%%%%%%%%%%%%%%%%%%%%%%%%%%%%%%%
%%
%%   dvips -P cmz -t letter -o <file>.ps <file>.dvi
%%
%%%%%%%%%%%%%%%%%%%%%%%%%%%%%%%%%%%%%%%%%%%%%%%%%%%%%
%%   for arXiv submission:
%% 1: put all separate .tex files into one .tex file
%% 2: for urls in the bib with proper line break, use:
%%\usepackage{url}
%%\def\UrlBreaks{\do\/\do-}
%%\usepackage[breaklinks]{hyperref}
%%\usepackage{breakurl}
\end{abstract}

%%introduction
%\input{introduction}
\section{Introduction}\label{Sec:Introduction}
In the academic community, it is customary to get a quick impression
of an author's research from simple statistics about his/her
publications. Such statistics include paper count, citations of
papers, h-index and various other indices for counting papers and
citations. Several services, such as ISI, Scopus, Google Scholar,
CiteSeerX~\cite{citeseer}, DBLP~\cite{dblp} and
Microsoft~\cite{libra}, facilitate the retrieval of these statistics
by maintaining databases indexing the metadata of academic
publications. These databases are usually proprietary and the
information users can retrieve, sometimes on a paid basis, is
limited to what these services choose to provide.

In recent years, some of these service providers \cite{dblp,libra}
are making the database more publically accessible and are starting
to provide additional information users can query (this is specially
the case with Libra). This allows us to study the author community
as a social network, analyzing not only the statistics about papers
published by an author, individually at a time, but also an author's
choice and extent in \emph{connecting} to other authors
(co-authoring) and an author's \emph{influence} on other authors.
Since citation is a \emph{slow} indicator for evaluating an author's
standing, we can also design metrics to measure an author's
\emph{exposure} in her research community, to estimate his/her
future influence and connections in research.

Our approach is to design various social network types of metrics to
measure the traits defined above. Since there is no ground-truth for
validation, we justify our designs by the following methods: (1)
Compare top ranked authors to those receiving awards for qualities
similar to what we try to measure, e.g. influence; (2) Use
similarity study to ensure any new metric can measure something
different from that is indicated by other well-established metrics
already; (3) Undertake case-studies of those authors scoring very
differently under different metrics, in domains we are familiar
with; (4) Let colleagues use our experimental website
(\url{http://pubstat.org}) and get their feedback on its usefulness.

Our conclusion is that several of the metrics we designed, namely
\emph{Influence}, \emph{Connections} and \emph{Exposure}, can
provide different rankings of authors, and together with Citation
Count can give a fuller picture about authors.

%% newly added start
According to the author ranking results, combined with additional
information on author-institution relationships, we further studied
and designed approaches for conducting author-based institution
ranking for each of the various metrics as well as the subject
domains.
%% newly added end

In the rest of the paper, we first describe briefly the available
dataset. We then describe the metrics we studied and the ranking
services we built. Next we evaluate our metrics and ranking methods
using the approach described above. We finish by discussing related
works and our conclusions.

%dataset
%\input{dataset}
%%section: dataset
\section{Data}
\begin{figure}
    \centering
    \includegraphics[width=2.6in]{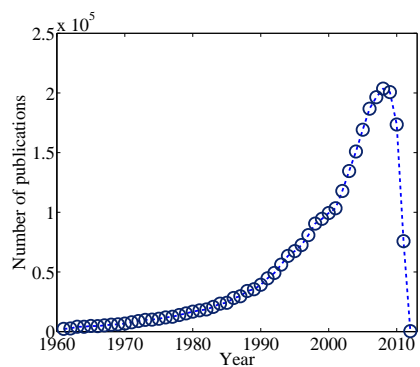}
    \caption{The number of papers in the Libra CS domain changing with time.}
    \label{Fig:LibraPaperYear}
\end{figure}

Our data is collected from the Microsoft Libra public API. The Libra
data has an object-level organization~\cite{nie2007object}, which is
very helpful. The object type includes: author, paper, conference
venue, institution and so on. Each type of object possess general
properties such as a unique identifier, name and relationship to
other objects. For example, if the object is a paper, then its
properties include publication year, authorship and citations. In
fact, Libra has maintained a huge amount of data in a very wide
range of research fields (15) and, for each field, it further
categorizes the papers to belong to domains in that field. The data
set we obtained for experimental purposes was for the Computer
Science field, which included 24 domains.
Table~\ref{Tab:DataSetDomain} lists the name, the number of authors
and the number of papers in the domains. Since each author may
publish papers in different domains, the sum of authors in all
domains is significantly greater than the number of unique authors
(941733).  The number of papers in the database (3347795) is
actually significantly greater than the sum from all domains
(2449673). This is because many papers were not classified or had
missing information. Another fact we needed to consider was that an
increasing proportion of these papers were published in more recent
years, as shown in Fig.~\ref{Fig:LibraPaperYear}. This has some
ramifications for our analysis, as we discuss in the latter part of
this paper. Despite the misgivings about the dataset we make many
interesting observations.

\begin{table}[htb]
\centering \caption{The basic information of the Libra dataset we
use: domain name, the number of authors and the number of papers in
each of the 24 domains of the Computer Science Area.}
\label{Tab:DataSetDomain}
\begin{tabular}{|l|c|c|}
\hline Domain Name                              & \#Authors & \#Papers \\
\hline Algorithms and Theory                    & 96748     & 270601 \\
\hline Security and Privacy                     & 33910     & 61957 \\
\hline Hardware and Architecture                & 81021     & 150151 \\
\hline Software Engineering                     & 85938     & 174893 \\
\hline Artificial Intelligence                  & 186976    & 325109 \\
\hline Machine Learning and Pattern Recognition & 66839     & 108234 \\
\hline Data Mining                              & 50958     & 67485 \\
\hline Information Retrieval                    & 30038     & 51075 \\
\hline Natural Language and Speech              & 86670     & 220227 \\
\hline Graphics                                 & 36548     & 59880 \\
\hline Computer Vision                          & 44969     & 60806 \\
\hline Human-Computer Interaction               & 51548     & 79909 \\
\hline Multimedia                               & 59277     & 80618 \\
\hline Network and Communications               & 138096    & 235297 \\
\hline World Wide Web                           & 25098     & 35861 \\
\hline Distributed and Parallel Computing       & 69592     & 117836 \\
\hline Operating System                         & 18167     & 25395 \\
\hline Databases                                & 74125     & 142421 \\
\hline Real-Time and Embedded System            & 21965     & 33098 \\
\hline Simulation                               & 18083     & 27678 \\
\hline Bioinformatics and Computational Biology & 48729     & 55491 \\
\hline Scientific Computing                     & 103982    & 183878 \\
\hline Computer Education                       & 29420     & 49125 \\
\hline Programming Languages                    & 33229     & 70561 \\
\hline Computer Science Overall (24 domains)    & 941733    & 2449673 \\
\hline Computer Science Total Involved          & 1175052   & 3347795 \\
\hline
\end{tabular}
\end{table}

%%%%%%%%%%%%%%%%%%%%%%%%%%%%%%%%%%%%%%%
%metrics
%\input{metrics}
%%metrics section
\section{Metrics and Ranking Methods}\label{Sec:metrics}
\subsection{Metrics}
\begin{figure}[htb]
    \centering
    \includegraphics[width=3.6in]{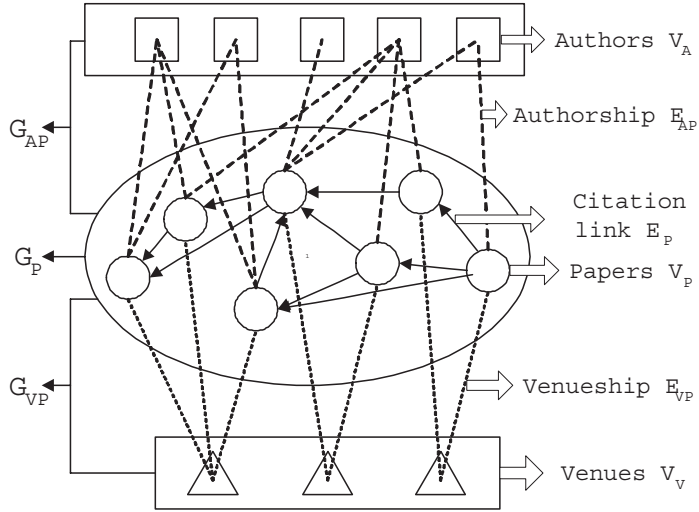}
    \caption{Underlay network topology of authors, venues and papers.}
    \label{FigTriPartite}
\end{figure}

All the metrics we studied can be defined by considering three types
of object: (a) papers, (b) authors and (c) venues. The relationships
between these objects are captured by the following networks
(graphs):
\begin{enumerate}
\item [a)] paper citation network, denoted by $G_P = (V_P, E_P)$, where $V_P$ is
the set of papers and $E_P$ is the set of citations from one paper
to another.\\
\item[b)] authorship bipartite network, denoted by $G_{AP} = (V_A \cup V_P,
E_{AP})$, where $V_A$ is the set of authors and edges in the set
$E_{AP}$ link each paper to its authors (authorship) and
symmetrically each author to his/her publications (ownership).\\
\item[c)] venueship bipartite network, denoted by $G_{VP} = (V_V \cup V_P, E_{VP})$,
where $V_V$ is the set of venues and the edges in $E_{VP}$ connect
each paper to its publishing venue. Topologically, $G_{VP}$ is
similar to $G_{AP}$. The main difference is that each paper can have
multiple authors while it can only be published in one venue.
\end{enumerate}
Fig.~\ref{FigTriPartite} shows the super-graph $G=(V,E)$ combining
all three networks together. In this case, $V = (V_P \cup V_A \cup
V_V)$ and $E = (E_P \cup E_{AP} \cup E_{VP})$.
%tripartite_example.eps
We also denote $n_A = |V_A|, n_V = |V_V|$ and $n_P = |V_P|$ as the
number of authors, venues and papers, respectively.

We grouped the metrics we defined into three categories. A metric
may be a simple count, such as citation count, or a value derived
iteratively using a PageRank-like algorithm.
\begin{enumerate}
  \item Paper based - In this case, each paper has a value defined by a metric.
  The value is distributed to the paper's authors in a way also determined by the
  metric.
  For this category, we studied three metrics: Citation count ({\bf CC}), Balanced citation
  count ({\bf BCC}) and Citation value ({\bf CV}). For CC and BCC, the paper's value was
  simply the citation count, which is well-defined.  In CC's case, each co-author
  received the citation count whereas in BCC's case, each co-author only received an
  equal fraction of the citation count. For CV, it was computed iteratively based on
  the citation graph $G_P$ and distributed to the co-authors in equal
  fractions.\\
  \item Author based - These metrics were computed based on author-to-author
  relationships directly. In this category, we studied three metrics:
  {\bf Influence}, {\bf Followers} and {\bf Connections}.
  All three were computed iteratively. For Influence, the author-to-author relationship
  was derived from the citation graph $G_P$ and authorship graph $G_{AP}$.
  Every time author $i$ cites author $j$'s paper, author $i$'s Influence was distributed to
  author $j$, split among the co-authors of $j$.
  For Followers, the author-to-author relationship was also derived from the citation
  graph, but depended on whether author $i$ cited author $j$ instead of how many times. If
  author $i$ cited author $j$, author $i$'s Follower value was distributed to author $j$
  without splitting among author $j$'s co-authors (which could be different for different
  papers). The author-to-author relationship for Connections was defined only based on
  the authorship graph $G_{AP}$. If author $i$ had co-authored a paper with author $j$,
  then author $i$'s Connections value was distributed to author $j$ and \emph{vice versa}.
  Note, another variation of Connections could also be defined so that every time author $i$
  co-authored with author $j$, they exchanged their Connections
  value.\\
  \item Author and venue based - In this category, we defined only one metric:
  {\bf Exposure}. This metric was computed by iterating on authors and venues together.
  It is easiest to think of venues also as a kind of author, thus we had an
  enlarged author set $V_A \cup V_V$. The author-to-author relationship was
  defined in the same way as Influence; so was the relationship for venue-to-venue.
  The author-to-venue and venue-to-author relationships were defined intuitively as
  follows: each time an author $i$ wrote a paper published in venue $k$, author $i$
  distributed his/her influence to venue $k$; similarly, each time a venue $k$ published
  a paper co-authored by $i$, author $i$ shared a fraction of venue $k$'s influence
  with $i$'s co-authors for that paper.
\end{enumerate}

Note, all these (7) metrics were defined so as to assign a value to
each author, to indicate some characteristics of that author. Since
citation count (CC) could be inflated by a large number of
co-authored papers, BCC and CV were alternative computations to
assign citation credits to authors. The metrics Influence and
Followers are intended to characterize an author's influence and
impact on other authors. The metric Connections is used to measure
an author's reach in the co-authorship network. Finally, Exposure is
intended to bring in the impact of the venues to help characterize
an author's potential influence that may not be reflected by
citations if the author's papers were relatively recent.

For a precise definition of the above metrics, it is necessary to
explain the PageRank algorithm. A brief treatment of PageRank and
the metrics definition by equations are included in the Appendix.

\subsection{Ranking}
Given the metrics we defined, we computed for each author his/her
ranking for each metric. An example of an author "J Smith" (with the
actual name anonymized) returned by our web service is listed in
Table~\ref{Tab:DataSetExample}:

\begin{table}[htb]
\centering \caption{An example of the different metric results
returned by our web service in the ``Network and Communications''
domain with actual author name anonymized.}
\label{Tab:DataSetExample}
\begin{tabular}{|c|c|c|c|c|c|c|c|c|}
\hline Value Type & Author &CC &BCC &CV &Inf &Fol &Con &Exp\\
%
%\hline J Smith &4919 &2659 &2957 &4214 &7431 &1885 &1756 \\
\hline Rank     & J Smith &4786 &2483 &2996 &4100 &7647 &2820 &1805 \\
%
%\hline RankPer($\%$)  & J Smith &$3.47$ &$1.8$ &$2.17$ &$2.97$ &$5.54$ &$2.04$ &$1.31$ \\
\hline RankPer  & J Smith &$3.47\%$ &$1.8\%$ &$2.17\%$ &$2.97\%$ &$5.54\%$ &$2.04\%$ &$1.31\%$ \\
%
%\hline  Cum. Value($\%$) &J Smith &72.25 &63.66 &58.45 &56.49 &59.51 &18.15 &26.91 \\
\hline  CumValue &J Smith &72.25\% &63.66\% &58.45\% &56.49\% &59.51\% &18.15\% &26.91\% \\
\hline
\end{tabular}
\end{table}

Actually, this ranking is for a specific domain (``Network and
Communications'') which has close to 138K authors in our database.
So this author is ranked well within the top ten percentile of this
domain he/she works in. In order to give this information, we also
allow the user to view the ranking in terms of percentile (denoted
by RankPer, the 3rd row in Table~\ref{Tab:DataSetExample}).

A third choice is to view the ranking information in terms of the
cumulative value of contribution by authors ranked ahead of the
target author (denoted by CumValue, the 4th row in
Table~\ref{Tab:DataSetExample}).

Finally, we considered it more appropriate to use a coarse
granularity for such ranking information (especially applied in the
case study in a later section).
There were two possible ways: (1) based on cumulative value of
contribution; (2) based on rank percentile.

\paragraph{Contribution based letter grading:}
for this purpose, we decided to divide the cumulative value range
into five fixed intervals, and assign letter grade ABCDE as ranks.
Lacking any better way to calibrate the partitioning, we simply used
20\%, 40\%, 60\% and 80\% as the thresholds. In this view, the above
example becomes (Table~\ref{Tab:DataSetExampleContriLetter}):
\begin{table}[htb]
\centering \caption{Example of the contribution based letter grades
for each metric, where A:$(0-20\%)$, B:$(20\%-40\%)$,
C:$(40\%-60\%)$, D:$(60\%-80\%)$ and E:$(80\%-100\%)$}.
\label{Tab:DataSetExampleContriLetter}
\begin{tabular}{|c|c|c|c|c|c|c|c|c|}
\hline  Value Type & Author &CC &BCC &CV &Inf &Fol &Con &Exp\\
%\hline  Cum. Value($\%$) &J Smith &72.25 &63.66 &58.45 &56.49 &59.51 &18.15 &26.91 \\
\hline  CumValue &J Smith &72.25\% &63.66\% &58.45\% &56.49\% &59.51\% &18.15\% &26.91\% \\
\hline  Contri. Letter   &J Smith &D  &D  &C  &C  &D  &A  &B \\
\hline
\end{tabular}
\end{table}

\begin{figure}
    \centering
    \includegraphics[width=2.4in]{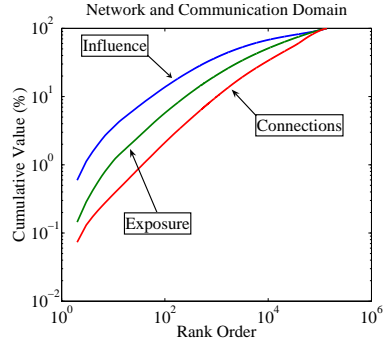}
    \caption{The loglog results of rank orders versus cumulative values of three metrics: Influence, Connections and Exposure.}
    \label{LogLogCumInfRank}
\end{figure}

For most metrics, the distribution of contribution by authors
ordered according to ranking follows Pareto-like distribution. For
example, Fig.~\ref{LogLogCumInfRank} shows the relationship between
the rank order to the cumulative value of three metrics, Influence,
Connections and Exposure, using a loglog plot.

So out of over 138K authors, the distribution of ABCDE for the
different metrics are listed in
Table~\ref{Tab:DataSetExampleContriDistri}.
\begin{table}[htb]
\centering \caption{The distribution of contribution based letter
assignment for different metrics of around 138k authors in ``Network
and Communications'' domain of Libra
dataset.}\label{Tab:DataSetExampleContriDistri}
\begin{tabular}{|l|c|c|c|c|c|c|c|}
\hline      &CC     &BCC    &CV     &Inf    &Fol    &Con    &Exp\\
\hline A    &156    &148    &179    &214    &485    &3386   &940 \\
\hline B    &558    &513    &752    &994    &1764   &11516  &3978 \\
\hline C    &1629   &1469   &2366   &4134   &5646   &32653  &12251 \\
\hline D    &5550   &5059   &9012   &25916  &26705  &20866  &31962 \\
\hline E    &130203 &130907 &125787 &106838 &103496 &69675  &88965 \\
\hline
\end{tabular}
\end{table}

\paragraph{Rank Percentile based letter grading:}
An alternative way of letter assignment was based on rank
percentile. Since the cumulative curves of the metrics show the
power-law property, we thus proposed the power-based thresholds
$(\alpha^4, \alpha^3, \alpha^2, \alpha)$ to assign letters according
to the rank percentile, where parameter $\alpha\in(0,1)$ controled
the skewness of the assignment results.
Table~\ref{Tab:DataSetExampleRankPerAssign} illustrates the letter
assignment results when we set $\alpha = 0.25$ for the experimental
web site.
\begin{table}[htb]
\centering \caption{The illustration of power-based letter
assignment according to the rank percentile with parameter $\alpha
\in (0, 1)$.}\label{Tab:DataSetExampleRankPerAssign}
\begin{tabular}{|r|c|c|}
\hline      &Rank Percentile        & $\alpha = 0.25$\\
\hline A    &$(0-\alpha^4)$         & $(0-0.39\%)$\\
\hline B    &$(\alpha^4-\alpha^3)$  & $(0.39\%-1.56\%)$\\
\hline C    &$(\alpha^3-\alpha^2)$  & $(1.56\%-6.25\%)$\\
\hline D    &$(\alpha^2-\alpha)$    & $(6.25\%-25\%)$\\
\hline E    &$(\alpha-1)$           & $(25\%-100\%)$\\
\hline
\end{tabular}
\end{table}

The letter grades according to the rank percentile on the ``J
Smith'' example are listed in
Table~\ref{Tab:DataSetExampleRankPerLetter}.
%% and the distribution of
%%ABCDE for the different metrics are listed in
%%Table~\ref{Tab:DataSetExampleRankPerDistri}.
\begin{table}[htb]
\centering \caption{Letter grades for each metric by power-based
assignment according to the rank percentile on the example ``J
Smith'', where $\alpha = 0.25$.}
\label{Tab:DataSetExampleRankPerLetter}
\begin{tabular}{|c|c|c|c|c|c|c|c|c|}
\hline  Value Type & Author &CC &BCC &CV &Inf &Fol &Con &Exp\\
%\hline  Cum. Value($\%$) &J Smith &72.25 &63.66 &58.45 &56.49 &59.51 &18.15 &26.91 \\
%\hline RankPer($\%$)  & J Smith &$3.47$ &$1.8$ &$2.17$ &$2.97$ &$5.54$ &$2.04$ &$1.31$ \\
\hline RankPer  & J Smith &$3.47\%$ &$1.8\%$ &$2.17\%$ &$2.97\%$ &$5.54\%$ &$2.04\%$ &$1.31\%$ \\
\hline RankPer Letter &J Smith &C  &C  &C  &C  &C  &C  &B \\
\hline
\end{tabular}
\end{table}

It remains an open problem of how to find the best way of letter
grades assignment, which we consider to be future work. We briefly
discuss the pros and cons of the two letter assignment methods
proposed by us, contribution based vs. rank percentile based.

One notable difference was the metric-dependency of the letter count
distribution. By definition, rank percentile based letter grading
results in a consistent letter count distribution among different
metrics (hence independent of metrics). However, it varies a lot
among different metrics for letter count distribution generated by
contribution based letter grading. For example, as shown in
Table~\ref{Tab:DataSetExampleContriDistri}, there were 156 ``A''s
for the Citation Count (CC) metric when 3386 ``A''s for the
Connection (Con) metric. This was caused by the different skewness
in the value distribution of authors' contribution for various
metrics, which could also be inferred from the cumulative curves
shown in Fig.~\ref{LogLogCumInfRank}.

On the other hand, any change in the total number of authors in a
research domain (e.g. community expansion or rapid development)
unavoidably affects the letter count distribution generated by rank
percentile based letter grading, but it has very limited effects on
the contribution based letter assignment results when the value
distribution is very skewed (e.g. Influence etc.).

Later on, unless otherwise noted, we only show the letter assignment
results by percentile based grading for space saving and fair
comparison among various metrics.

\subsection{Domain-specific vs Overall Ranking}

As mentioned, the above example is the ranking for an author in a
specific domain. Usually, an author works in several domains. Our
web service shows the author's rankings in all the domains, as well
as an overall score for his/her subject field (in this case
``Computer Science''). The letter grades of the example ``J Smith''
are listed in Table~\ref{Tab:DataSetExampleDomains}).
\begin{table}[htb]
\centering \caption{Letter grades of each metric in several involved
domains of the example ``J Smith''.}
\label{Tab:DataSetExampleDomains}
\begin{tabular}{|l|c|c|c|c|c|c|c|}
\hline Domain &CC &BCC &CV &Inf &Fol &Con &Exp\\
\hline Net\&Comm &C&C&C&C&C&C&B \\
\hline Sec\&Priv &D&D&D&D&E&E&D \\
\hline Overall   &C&C&C&C&C&C&B \\
\hline
\end{tabular}
\end{table}

This allows the person to be compared to others in his/her domain,
as well as comparing him/her to a bigger set of people in a subject
field.

The way to compute the overall score is difficult. We used the
straightforward way of merging all the domains into one big domain
and compared the results. This was more computationally demanding.
Another possible way would be to add up the authors ranking in each
domain normalized by the size of each domain. The trade-off of
different ways for computing the overall is something still under
study.

\subsection{Comparing Rankings}

In our experimental web site, we have implemented different ways for
authors to be compared. First of all, authors in the same domain can
be looked up in ranking order, according to any metric. So it would
be easy to look up top-ranked people according to one's favourite
metric, whether it was Influence, Connections, or Exposure. This is
often helpful.

Second, we allow authors in the same institution to be looked up in
ranking order, for a specific domain, or according to overall
ranking. This would be useful in getting a feel as to how strong a
particular institute was in a particular domain. It is also the
rough way we justify our assignment of ABCDE to authors in different
cumulative value percentiles or rank percentiles.

We also allow users to search for individual authors and keep them
in a list for head-to-head comparison. This could be helpful for
many different purposes. For example, we could use this method to
collect a list of authors for a case study (see next section).

We have also implemented various other features. For example, it
would be possible to look at all rankings if we excluded
self-citations. Basically, for each common query users find useful,
we could implement it as an additional feature.

\subsection{Author-based Institution Rankings}

With the additional information of author-institution relationships,
we can further provide institution rankings based on authors'
ranking results. When ranking institutions, we used two
granularities:
\begin{enumerate}
\item[(1)] We only count the number of authors assigned with ``A'';
\item[(2)] We compute a total score, counting ``A''=1, ``B''=0.5,
``C''=0.25, and ``D''=``E''=0.
\end{enumerate}

For ranking authors, there are a number of various metrics (e.g.,
Influence, Connections, Exposure, etc.), two types of letter
assignment (contribution based vs. rank percentile based) and the
domain-specificity (e.g., 24 domains listed in
Table~\ref{Tab:DataSetDomain}), therefore the institution ranking
automatically inherits these features.

%
%\section{Evaluation and Validation}
%\input{validation}
%%evaluation and validation
\section{Evaluation and Validation}\label{Sec:Validation}

\subsection{Ranking Award Recipients}

One way to justify our new metrics is to look at award recipients.
In the computer science domain, the most prestigious award is the
Turing Award.  Since we are more familiar with the Network and
Communications domain, we also looked at the ACM Sigcomm Award
recipients.  The results are shown in the following two tables
(Table~\ref{Tab:Turing} and Table~\ref{Tab:Sigcomm}).

%%%%%%%Award samples(Sigcomm and turing award)
%%%%%turing award
\begin{table*}
\centering \caption{Rankings received by Turing Award Recipients}
\begin{tabular}{|l|c|c|c|c|c|c|c|c|c|c|c|}
\hline Year &Awardee &In/All &h &CC &BCC &CV &Inf &Fol &Con &Exp &Aff \\
\hline 1966 &Alan J. Perlis         &31/47  &9&B&B&B&A&A&C&B      &Yale\\
\hline 1967 &Maurice V. Wilkes      &50/100 &11&B&B&A&A&A&D&A      &Cambridge\\
\hline 1968 &Richard W. Hamming     &9/29   &8&B&A&A&A&A&E&B      &Naval Postgraduate Sch.\\
\hline 1969 &Marvin Minsky          &50/79  &20&A&A&A&A&A&D&A      &MIT\\
\hline 1970 &James H. Wilkinson     &23/48  &9&B&A&A&A&A&D&A      &Nat. Physical Lab, UK\\
\hline 1971 &John McCarthy          &117/209&29&A&A&A&A&A&B&A      &Princeton\\
\hline 1972 &Edsger W. Dijkstra     &84/121 &28&A&A&A&A&A&C&A      &UT Austin\\
\hline 1973 &Charles W. Bachman     &18/25  &7&C&B&B&A&B&C&B      &Bachman Info Systems\\
\hline 1974 &Donald E. Knuth        &179/241&40&A&A&A&A&A&B&A      &Stanford\\
\hline 1975 &Allen Newell           &139/192&33&A&A&A&A&A&A&A      &Carnegie Mellon Univ\\
            &Herbert Simon          &140/398&33&A&A&A&A&A&B&A      &Illinois Institute of Tech\\
\hline 1976 &Michael O. Rabin       &68/81  &28&A&A&A&A&A&C&A      &Columbia\\
            &Dana Stewart Scott     &48/71  &21&A&A&A&A&A&C&A      &Carnegie Mellon Univ.\\
\hline 1977 &John W. Backus         &32/73  &11&A&A&A&A&A&C&A      &IBM\\
\hline 1978 &Robert W. Floyd        &36/46  &16&A&A&A&A&A&D&A      &Illinois Institute of Tech\\
\hline 1979 &Kenneth E. Iverson     &43/70  &10&C&B&B&A&A&C&A      &IBM\\
\hline 1980 &C. A. R. Hoare         &198/249&41&A&A&A&A&A&B&A      &Microsoft Research\\
\hline 1981 &Edgar Frank Codd       &27/32  &15&A&A&A&A&A&D&A      &IBM\\
\hline 1982 &Stephen A. Cook        &127/138&32&A&A&A&A&A&B&A      &Univ of Michigan\\
\hline 1983 &Ken Thompson           &26/51  &13&A&A&A&A&A&C&A      &Google\\
            &Dennis M. Ritchie      &29/37  &15&A&A&A&A&A&D&A      &Bell Labs\\
\hline 1984 &Niklaus Emil Wirth     &110/144&30&A&A&A&A&A&D&A      &Xerox PARC\\
\hline 1985 &Richard Manning Karp   &277/325&61&A&A&A&A&A&A&A      &IBM\\ %
\hline 1986 &John Edward Hopcroft   &147/176&39&A&A&A&A&A&B&A      &Stanford\\
            &Robert Endre Tarjan    &338/362&72&A&A&A&A&A&A&A      &Hewlett-Packard\\
\hline 1987 &John Cocke             &45/52  &20&A&A&A&A&A&C&A      &IBM\\
\hline 1988 &Ivan E. Sutherland     &57/63  &21&A&A&A&A&A&B&A      &Portland State Univ\\
\hline 1989 &William Morton Kahan   &32/39  &11&C&C&B&B&B&B&C      &UC Berkeley\\ %
\hline 1990 &Fernando Jose Corbato  &7/13   &5&C&C&B&B&A&D&C      &MIT\\
\hline 1991 &Robin Milner           &143/172&47&A&A&A&A&A&B&A      &Cambridge\\
\hline 1992 &Butler W. Lampson      &116/140&36&A&A&A&A&A&B&A      &MIT\\
\hline 1993 &Juris Hartmanis        &115/140&25&A&A&A&A&A&B&A      &Cornell\\
            &Richard Edwin Stearns  &77/89  &20&A&A&A&A&A&B&A      &NY Univ at Albany\\
\hline 1994 &Edward A. Feigenbaum   &37/58  &14&B&B&A&A&A&C&A      &Stanford\\
            &Raj Reddy              &70/99  &14&B&B&A&A&A&B&A      &Cargegie Mellon Univ\\
\hline 1995 &Manuel Blum            &100/112&33&A&A&A&A&A&B&A      &Carnegie Mellon Univ\\
\hline 1996 &Amir Pnueli            &331/371&62&A&A&A&A&A&A&A      &New York Univ\\
\hline 1997 &Douglas C. Engelbart   &23/31  &14&B&A&A&A&A&C&A      &Doug Engelbart Institute\\
\hline 1998 &Jim Gray               &217/293&46&A&A&A&A&A&A&A      &Microsoft Research\\
\hline 1999 &Fred Brooks            &77/112 &21&A&A&A&A&A&A&A      &UNC\\
\hline 2000 &Andrew Chi-chih Yao    &159/183&35&A&A&A&A&A&B&A      &Tsinghua Univ\\
\hline 2001 &Ole-johan Dahl         &32/39  &13&B&B&A&A&A&C&B      &Univ of Oslo\\
            &Kristen Nygaard        &35/43  &14&B&B&A&A&A&C&B      &Univ of Oslo\\
\hline 2002 &Ronald L. Rivest       &226/267&52&A&A&A&A&A&A&A      &MIT\\
            &Adi Shamir             &186/206&46&A&A&A&A&A&A&A      &Weizmann Institute\\
            &Leonard Max Adleman    &72/86  &27&A&A&A&A&A&B&A      &MIT\\
\hline 2003 &Alan Curtis Kay        &22/33  &7&B&B&B&B&B&C&B      &Hewlett-Packard Labs\\
\hline 2004 &Vinton Gray Cerf       &39/56  &10&B&B&B&A&A&B&B      &Google\\
            &Robert Elliot Kahn     &15/23  &9&C&C&A&A&A&C&B      &CNRI\\
\hline 2005 &Peter Naur             &40/137 &7&C&B&A&A&A&D&A      &Univ of Copenhagen\\
\hline 2006 &Frances E. Allen       &26/37  &14&B&B&A&A&A&C&B      &IBM\\
\hline 2007 &Edmund Clarke          &333/370&63&A&A&A&A&A&A&A      &Carnegie Mellon Univ\\
            &E. Allen Emerson       &132/150&41&A&A&A&A&A&B&A      &UT Austin\\
            &Joseph Sifakis         &139/164&36&A&A&A&A&A&A&A      &CNRS\\
\hline 2008 &Barbara Liskov         &195/233&48&A&A&A&A&A&A&A      &MIT\\
\hline 2009 &Charles P. Thacker     &12/15  &7&B&B&B&A&A&C&C      &Microsoft\\
\hline 2010 &Leslie Valiant         &113/124&37&A&A&A&A&A&C&A      &Harvard Univ\\
\hline 2011 &Judea Pearl            &193/258&39&A&A&A&A&A&B&A      &UCLA\\
\hline 2012 &Shafi Goldwasser       &138/152&43&A&A&A&A&A&B&A      &Weizmann Institute\\
            &Silvio Micali          &165/173&46&A&A&A&A&A&B&A      &MIT\\
\hline
\end{tabular}
\label{Tab:Turing}
\end{table*}
%\newpage
%\newpage
%%%%%Sigcomm award
\begin{table*}
\centering \caption{Rankings received by Sigcomm Award recipients}
\begin{tabular}{|l|c|c|c|c|c|c|c|c|c|c|c|c|c|}
\hline Year &Awardee &In/All &h &CC &BCC &CV &Inf &Fol &Con &Exp &Aff.\\
\hline 1989 &Paul Baran             &1/7    &1&D&D&C&C&C&E&D      &RAND Corporation\\
\hline 1990 &Leonard Kleinrock      &173/233&31&A&A&A&A&A&A&A      &UCLA\\
            &David D. Clark         &43/80  &13&B&B&B&A&A&B&B      &MIT\\
\hline 1991 &Hubert Zimmermann      &10/17  &4&C&B&B&B&B&E&B      &Sun Microsystems\\
\hline 1992 &A. G. Fraser           &16/24  &6&C&C&B&A&B&E&B      &Fraser Research\\
\hline 1993 &Robert Elliot Kahn     &15/23  &9&C&C&A&A&A&C&B      &CNRI\\
\hline 1994 &Paul E. Green          &20/54  &7&C&B&B&B&B&C&B      &Tellabs\\
\hline 1995 &David J. Farber        &43/55  &13&B&B&B&B&A&B&B      &Carnegie Mellon Univ\\
\hline 1996 &Vinton Gray Cerf       &39/56  &10&B&B&B&A&A&B&B      &Google\\
\hline 1997 &Jonathan B. Postel     &73/92  &24&A&A&A&A&A&B&A      &Univ of Southern California\\
            &Louis Pouzin           &7/23   &3&D&C&B&B&B&E&C      &ITU\\
\hline 1998 &Lawrence G. Roberts    &14/16  &8&B&A&A&A&A&D&A      &Anagran Inc.\\
\hline 1999 &Peter T. Kirstein      &45/67  &7&C&C&B&B&B&C&B      &Univ College London\\
\hline 2000 &Andre A. S. Danthine   &31/44  &7&C&C&C&C&C&C&C      &Universit\'e de Li\'ege\\
\hline 2001 &Van Jacobson           &113/126&41&A&A&A&A&A&B&A      &Palo Alto Research Center\\
\hline 2002 &Scott J. Shenker       &413/481&88&A&A&A&A&A&A&A      &UC Berkeley\\
\hline 2003 &David Cheriton         &156/186&36&A&A&A&A&A&B&A      &Stanford\\
\hline 2004 &Simon Lam              &148/181&30&A&A&A&A&A&B&A      &UT Austin\\
\hline 2005 &Paul V. Mockapetris    &17/21  &8&B&A&A&A&A&E&A      &Nominum\\
\hline 2006 &Domenico Ferrari       &101/128&30&A&A&A&A&A&B&A      &UC Berkeley\\
\hline 2007 &Sally Floyd            &186/206&59&A&A&A&A&A&A&A      &ICSI\\
\hline 2008 &Donald F. Towsley      &618/725&65&A&A&A&A&A&A&A      &Univ of Massachusetts\\
\hline 2009 &Jon Crowcroft          &284/375&42&A&A&A&A&A&A&A      &Cambridge\\
\hline 2010 &Radia J. Perlman       &22/25  &11&B&B&A&A&B&D&B      &Intel\\
\hline 2011 &Vern Paxson            &182/212&54&A&A&A&A&A&A&A      &UC Berkeley\\
\hline 2012 &Nick W. Mckeown        &140/179&34&A&A&A&A&A&A&A      &Stanford\\
\hline
\end{tabular}
\label{Tab:Sigcomm}
\end{table*}
%\end{landscape}

In these tables, the two numbers in the third column (In/All) are
the number of papers we considered ``In Domain'' and used for
computing the ranking, and the total number of papers authored by
the author. In both these cases, it is clear that citation count is
not always a good measure, for these people obviously had tremendous
contribution and impact in their fields. The Citation Value metric
(CV) improved over CC and BCC. But Influence did much better - all
the Turing Award winners scored at least B.  For these top people in
their fields, the Followers metric was even more predictive. Though,
as we will discuss later, we find Influence and Followers quite
similar. Aside from trying to justify the Influence and Followers
metrics, we can also appreciate the additional information provided
by the Connections metric, in distinguishing those who tend to
collaborate more from those who tend to work alone.

Since Sigcomm is a more applied community, the CC and BCC metrics
performed even worse in comparison to Influence and Followers. This
is perhaps because the Sigcomm community publication venues are more
selective (hence have more influence). We will discuss the
differences between Influence, Followers and Exposure later.

\subsection{Similarity between proposed Metrics}
\begin{figure}[h]
    \centering
    \includegraphics[width=2.3in]{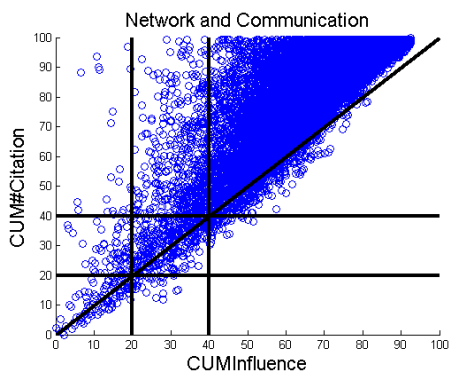}
    \caption{Comparison between Influence and Citation Count using cumulative value.}
    \label{CumInfCit}
\end{figure}

For our similarity study, we chose to plot the cumulative value
(essentially according to letter grades) of each author, for the two
comparable metrics. For example, we first compared Citation Count
(CC) with Influence as metrics. The former was the common metric
used in practice, and the latter was something we proposed. The
result is shown in Fig.~\ref{CumInfCit}.
The two vertical and horizontal lines give the boundaries separating
A and B from the rest of the ranks. Any author on the diagonal line
received exactly the same ranking from both metrics.  As we can see,
there is correlation between Influence and CC - those with high CC
ranking all have high Influence ranking as well. But the converse is
not true - those with high Influence ranking may not have high CC
ranking. This means we could use CC as a sufficient condition when
estimating someone's influence, but not a necessary condition. For
this reason, we consider Influence is sufficiently different than
CC, and should be considered as a complementary metric.

\begin{figure}[h]
    \centering
    \includegraphics[width=2.3in]{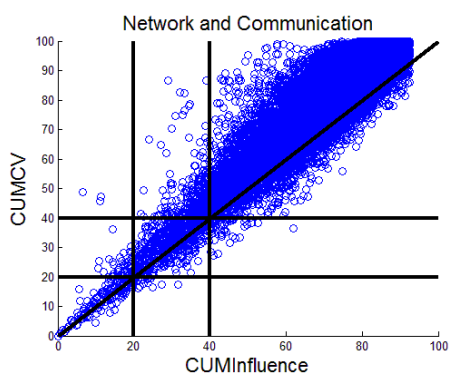}
    \caption{Comparison between Influence and Citation Value using cumulative value.}
    \label{CumInfPta}
\end{figure}

The Citation Value (CV) metric was designed to be an alternative to
CC. From our experience, an author's CV rank seems to be always
between its CC rank and Influence rank. Fig.~\ref{CumInfPta}
compares CV against Influence.
It is indeed similar to the comparison to CC, namely high CV implies
high Influence but not \emph{vice versa}. Thus, once we have CC and
Influence, there is no strong reason to keep CV as an additional
metric.

\begin{figure}[h]
    \centering
    \includegraphics[width=2.3in]{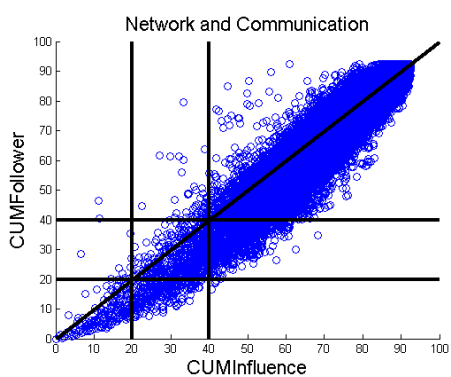}
    \caption{Comparison between Influence and Follower using cumulative value.}
    \label{CumInfFol}
\end{figure}

Now let us consider the Followers metric. As we observed in
considering the Followers and Influence ranks for the Award
recipients, those with a high influence rank tend to have even
higher Followers ranks. But for the majority of the authors, these
two ranks are very strongly correlated, and hence Followers seem to
add little additional value to the Influence metric (as shown in
Fig.~\ref{CumInfFol}).

As expected, the Connections metric had little correlation to any of
the other metrics. This is quite intuitive, so we have not included
any similarity plots to save space.

\begin{figure}[h]
    \centering
    \includegraphics[width=2.3in]{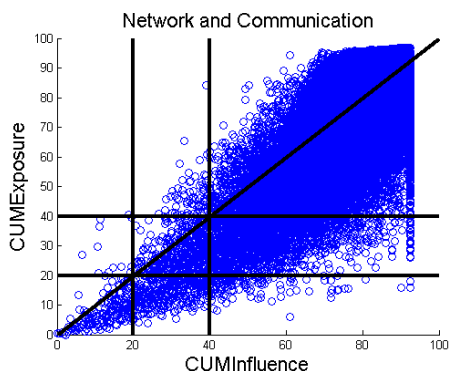}
    \caption{Comparison between Influence and Exposure using cumulative value.}
    \label{CumInfExp}
\end{figure}

Finally, we compared the Influence metric to the Exposure metric in
Fig.~\ref{CumInfExp}.
In this case,many authors with low Influence values may have much
higher ranks in Exposure. We suspect this is because this metric
successfully identifies authors who are very active in publishing in
high impact venues but have not had the time to build up their
influence. It is difficult to tell how true this is - so we selected
some real world examples for our case studies in a later
subsection.

\subsection{Similarity study with h-index}
Next we investigated the similarity between the newly proposed metrics to the well known h-index~\cite{hindexNature,hindex}.
\begin{figure}[h]
    \centering
    \includegraphics[width=2.3in]{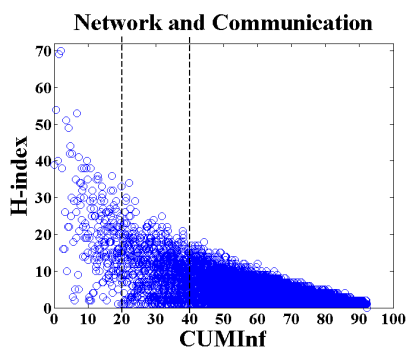}
    \caption{Comparison between cumulative Influence value and h-index.}
    \label{CumInfHindex}
\end{figure}

We first compared the Influence metric to the h-index in Fig.~\ref{CumInfHindex}.
It is similar to the correlation between CC and Influence,
i.e., those with high h-indices all have high Influence rankings as well. But the converse is
not true - those with high Influence rankings may not have high h-indices.
%This again indicates that Influence indeed provides complementary information than the
%widely used practical metrics like citation counts, h-index and so on.
%This gives additional insight for why influence may be a better
%metric to use instead of h-index.
This reinforces the belief that influence is a better metric to differentiate those authors with high h-indices.

\begin{figure}[h]
    \centering
    \includegraphics[width=2.3in]{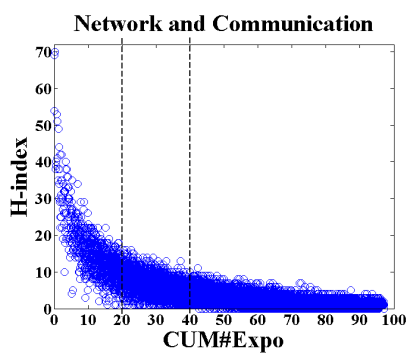}
    \caption{Comparison between cumulative Exposure value and h-index.}
    \label{CumExpHindex}
\end{figure}

Next we compared the Exposure metric to the h-index in Fig.~\ref{CumExpHindex}.
It shows again that high h-indices implies high Exposure rankings while the converse is not true.
%But there is a
A clear difference that more points are located at the bottom left
area, when comparing to Fig.~\ref{CumInfHindex}. This is consistent
with our suspicion that there exist many authors who are very active
in publishing in high impact venues but their h-index values have
not had enough time to accumulate. A similar argument was also
raised by~\cite{RefOnHindex}.

\begin{figure}[h]
    \centering
    \includegraphics[width=2.3in]{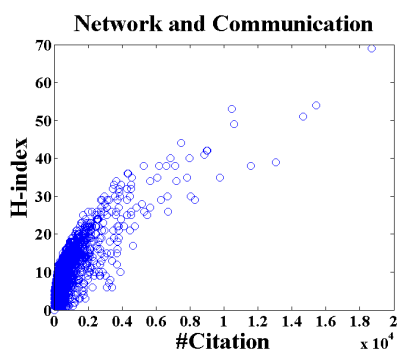}
    \caption{Comparison between total citation count and h-index.}
    \label{ccHindex}
\end{figure}

At last, we looked into the total citation count versus the h-index of each author in Fig.~\ref{ccHindex}.
As expected, the correlation between total citation counts and h-indices generally follows the square root law.
This comes from the definition of the h-index~\cite{hindex}.

\subsection{Case Studies}

From the above similarity study, we concluded that, out of the five
metrics based on iterative computation, i.e. CV, Influence,
Followers, Connections and Exposure, the first three are
sufficiently similar: we therefore chose to keep only Influence.
Influence, Connections and Exposure are sufficiently different from
each other, and from CC.

For case studies, we considered two cases: (a) authors with high
Influence but low Citation Count; and (b) authors with high Exposure
but low Influence. (a) was the reason for keeping Influence, and (b)
was the reason for keeping Exposure. We selected some such cases in
the Network and Communications domain and show them in
Table~\ref{Tab:HighInfLowCC} and Table~\ref{Tab:HighExpLowInf}.

%%%%%Case Study for Influence and Citation Count
%%%High Influence and Low Citation Count
\begin{table}
\centering \caption{Examples for High Influence and Low CC}
\label{Tab:HighInfLowCC}
\begin{tabular}{|l|c|c|}
\hline Author                   &Influence  &\#Citation \\
\hline Robert Elliot Kahn       &A&C \\
\hline J. M. Wozencraft         &A&C \\
\hline Jean-Jacques Werner      &A&C \\
\hline David G. Messerschmitt   &A&C \\
\hline Nathaniel S. Borenstein  &A&C \\
\hline James L. Massey          &A&C \\
\hline W. T. Webb               &A&C \\
\hline Takashi Fujio            &A&D \\
\hline Martin L. Shooman        &A&D \\
\hline Sedat Olcer              &A&D \\
\hline Massimo Marchiori        &A&D \\
\hline Roger A. Scantlebury     &A&D \\
\hline
\end{tabular}
\end{table}

%%%High Influence and Low Exposure
%\begin{table}
%\centering \caption{The examples for High Influence and Low
%Exposure}
%\begin{tabular}{|l|c|c|c|c|}
%\hline AuthorName &In/AllPub   &Influence  &Exposure &Affiliation\\
%\hline D. L. A. Barber (1964-1979) &3/9    &A  &C  &National
%Physical Laboratory\\ \hline
%\end{tabular}
%\end{table*}

%%%High Exposure and Low Influence
\begin{table}
\centering \caption{Examples for High Exposure and Low Influence}
\label{Tab:HighExpLowInf}
\begin{tabular}{|l|c|c|}
\hline Author                   &Influence  &Exposure \\
\hline Achille Pattavina        &C&A \\
\hline Herwig Bruneel           &C&A  \\
\hline Yigal Bejerano           &C&A  \\
\hline Torsten Braun            &C&A  \\
\hline Kenneth J. Turner        &C&A  \\
\hline Ioannis Stavrakakis      &C&A  \\
\hline Emilio Leonardi          &C&A  \\
\hline Luciano Lenzini          &C&A  \\
\hline Dmitri Loguinov          &C&A  \\
\hline Romano Fantacci          &C&A  \\
\hline Hossam S. Hassanein      &C&A  \\
\hline Azzedine Boukerche       &C&A  \\
\hline
\end{tabular}
\end{table}

\subsection{Relation of Ranking to Publication Years}
%Comparison between Year Info and metrics

\begin{figure}[h]
    \centering
    \includegraphics[width=2.3in]{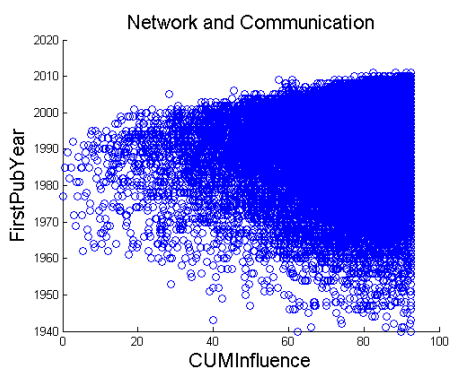}
    \caption{Comparison between Influence and the year of first publication.}
    \label{CumInfFirstYear}
\end{figure}

Finally, we were curious to find out the relationship between how an
author ranked and his/her first (or last) year of publication.
Fig.~\ref{CumInfFirstYear} plots the authors' Influence ranks
against their first year of publication.

It is worth noting that it takes time to build up Influence. Authors
ranked as A in Influence started publishing in the 1990s or earlier;
B authors started publishing in the early 2000s or earlier, and so
on (here is the contribution based letter assignment).

\begin{figure}[h]
    \centering
    \includegraphics[width=2.3in]{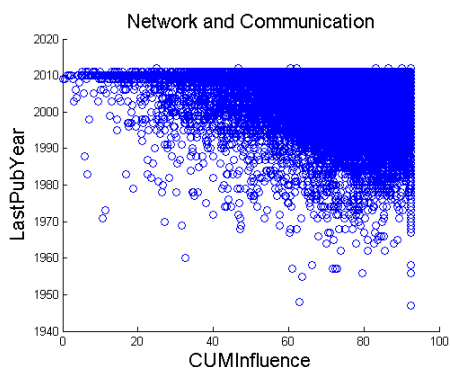}
    \caption{Comparison between Influence and the year of last publication.}
    \label{CumInfLastYear}
\end{figure}

\begin{figure}[h]
    \centering
    \includegraphics[width=2.3in]{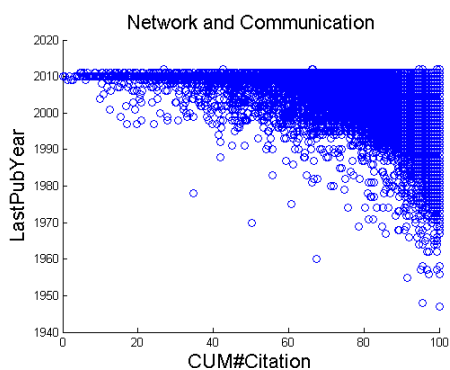}
    \caption{Comparison between Citation Count and the year of last publication.}
    \label{CumCitLastYear}
\end{figure}

\begin{figure}[h]
    \centering
    \includegraphics[width=2.3in]{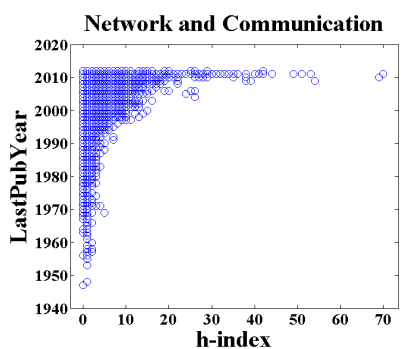}
    \caption{Comparison between h-index and the year of last publication.}
    \label{hinLastYear}
\end{figure}

Next, we plotted an author's last year of publication against
Influence (Fig.~\ref{CumInfLastYear}), Citation
Count (Fig.~\ref{CumCitLastYear}), and then against h-index
(Fig.~\ref{hinLastYear}), for comparison. Note, for
Citation Count and h-index, the high ranking people are mostly still active,
because we have been seeing paper and citation inflation over years.
For Influence, however, there is more \emph{memory}, in the sense
that more people who are no longer active also enjoy high Influence.
This is because an author's influence propagates, by definition of
the Influence metric.

%Comarison between Year Info and Citation Count
%\begin{figure}
%    \centering
%    \includegraphics[width=2.4in]{./epsFigures/CumCit_FPY2014.eps}
%    \caption{The comparison between Citation Count and the Year of authors first publication.}
%    \label{CumCitFirstYear}
%\end{figure}

%\newpage
\subsection{Author-based Institution Rankings}
In Table~\ref{Tab:AffRanking}, we illustrate the possibility of
institutional ranking according to authors' rankings in various
metrics. We selected 30 well-known universities and applied two
counting granularities on authors' letter grades of overall
``Computer Science'' rankings of three metrics, Citation Counts
(CC), Influence (Inf) and Exposure (Exp).
%%%%%%%Award samples(Sigcomm and turing award)
%%%%%turing award
\begin{table*}[thb]
\centering \caption{Illustration of Institution Rankings on 30
selected top universities of three metrics (\#Citations, Influence
and Exposure) at two counting granularities based on authors'
overall ranking in ``Computer Science'' domain.}
\label{Tab:AffRanking}
\begin{tabular}{|c|c|c|c|c|c|c|}
\hline
\multirow{2}{*}{Institution Name}
      & \multicolumn{3}{|c|}{Total Score Rank} & \multicolumn{3}{|c|}{\#A Rank}\\
      \cline{2-7} & CC & Inf & Exp & CC & Inf & Exp\\
\hline Massachusetts Institute of Technology    & 2 & 1 & 2 & 1 & 1 & 2 \\
\hline Carnegie Mellon University               & 1 & 2 & 1 & 2 & 2 & 1 \\
\hline Stanford University                      & 3 & 3 & 3 & 4 & 4 & 4 \\
\hline University of California Berkeley        & 4 & 4 & 4 & 3 & 3 & 3 \\
\hline University of Illinois Urbana Champaign  & 5 & 5 & 5 & 6 & 6 & 5 \\
\hline University of Southern California        & 6 & 6 & 7 & 5 & 5 & 6 \\
\hline Georgia Institute of Technology          & 7 & 6 & 6 & 8 & 7 & 8 \\
\hline University of California San Diego       & 11& 8 & 9 & 7 & 8 & 7 \\
\hline University of Washington                 & 10& 9 & 14& 10& 9 & 13 \\
\hline University of Maryland                   & 8 & 9 & 8 & 9 & 11& 8 \\
\hline University of California Los Angeles     & 12& 11& 11& 12& 10& 12 \\
\hline University of Texas Austin               & 9 & 11& 10& 11& 14& 10 \\
\hline University of Michigan                   & 13& 13& 11& 14& 14& 16 \\
\hline Cornell University                       & 15& 14& 15& 13& 11& 15\\
\hline University of Cambridge                  & 16& 15& 21& 17& 17& 22\\
\hline Columbia University                      & 17& 16& 19& 21& 20& 17\\
\hline University of Wisconsin Madison          & 20& 17& 28& 18& 18& 22\\
\hline University of Toronto                    & 18& 18& 16& 16& 16& 13\\
\hline The French National Institute for        & 14& 19& 11& 24& 26& 22\\
Research in Computer science and Control        &   &   &   &   &   &   \\
\hline University of Pennsylvania               & 22& 20& 27& 21& 21& 22\\
\hline Rutgers, The State University of New Jersey  & 23& 21& 22& 29& 18& 22\\
\hline Swiss Federal Institute of Technology Zurich & 18& 22& 25& 23& 26& 28\\
\hline Harvard University                       & 30& 23& 39& 33& 31& 35\\
\hline University of California Irvine          & 25& 24& 23& 19& 21& 20\\
\hline Purdue University                        & 21& 25& 18& 19& 31& 17\\
\hline University of Minnesota                  & 25& 25& 24& 27& 31& 22\\
\hline University of Massachusetts              & 24& 25& 26& 31& 36& 30\\
\hline Princeton University                     & 27& 28& 29& 14& 13& 17\\
\hline Technion Israel Institute of Technology  & 31& 29& 19& 24& 23& 10\\
\hline University of Edinburgh                  & 29& 30& 29& 31& 40& 35\\
\hline
\end{tabular}
\end{table*}

We found that the ranking results by different metrics were similar
at the institution level. The noise at the author ranking results
were cancelled out to a certain extent after they were aggregated
for scores. When we used the two granularities: (1) count the number
of authors assigned with ``A'' and (2) compute the total score,
counting ``A''=1, ``B''=0.5, ``C''=0.25 and ``D''=``E''=0, for
method (2), the size of an institution was influential; whereas for
method (1), smaller schools also had a chance to rank very high. For
example, in Table~\ref{Tab:AffRanking}, Princeton University was
ranked 28th by method (2), but 13th by only counting the number of
``A'' authors, i.e. by method (1).

\begin{figure*}[hbt]
    \centering
    \subfigure[CC]{
    \label{Fig:ccRPtmVccRPal}
    \includegraphics[width=1.4in]{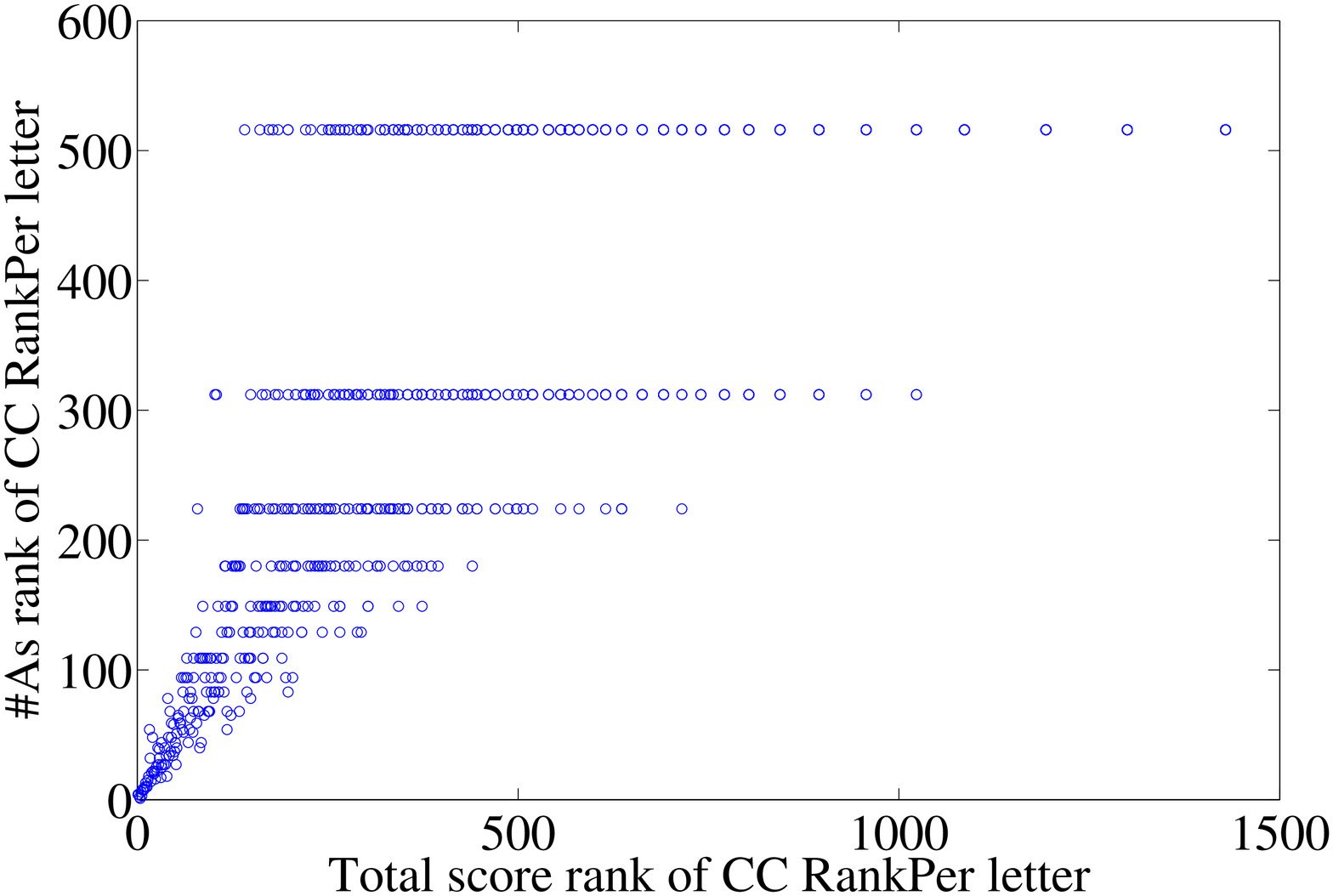}}
    \subfigure[Inf]{
    \label{Fig:infRPtmVinfRPal}
    \includegraphics[width=1.4in]{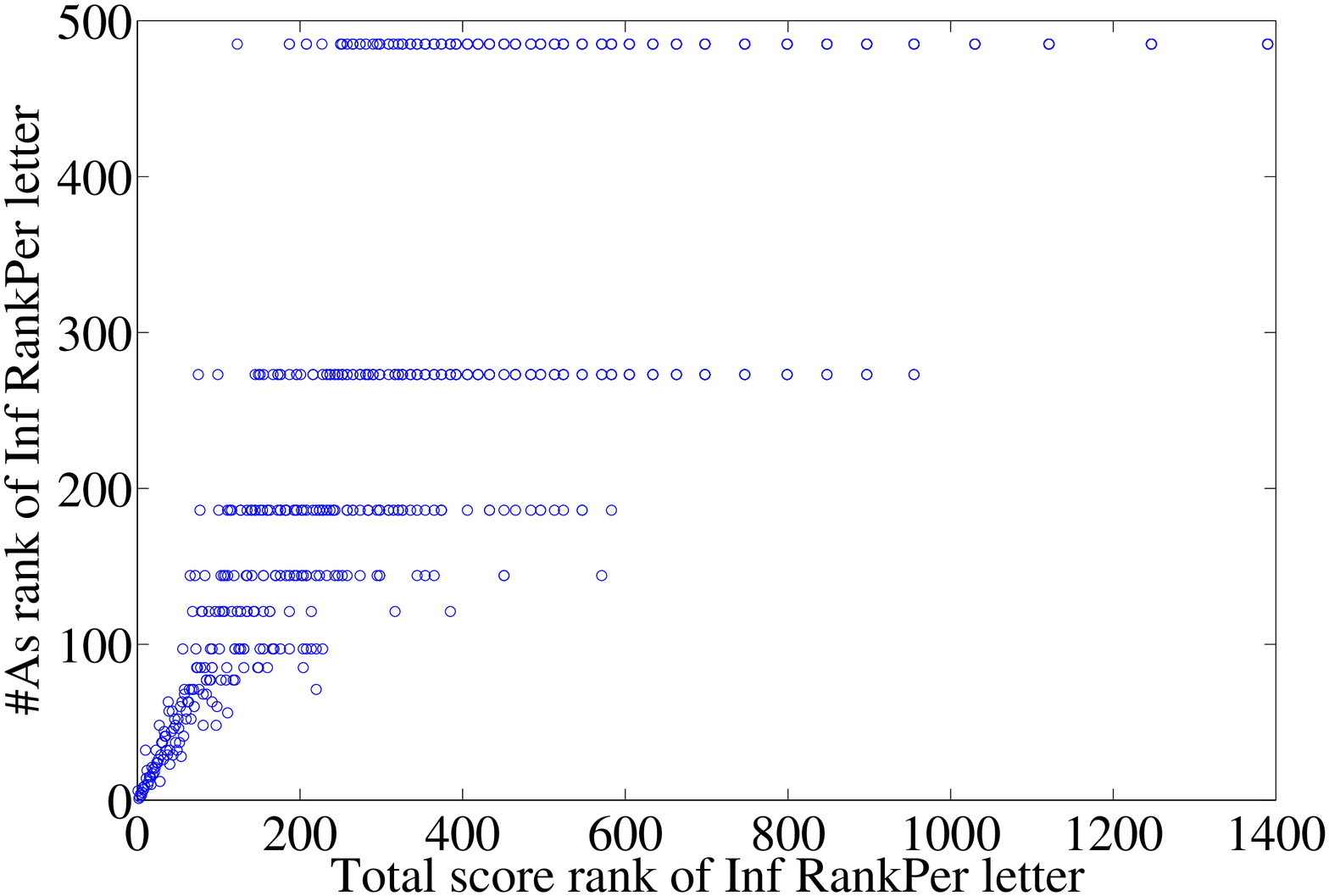}}
    \subfigure[Exp]{
    \label{Fig:avaRPtmVavaRPal}
    \includegraphics[width=1.4in]{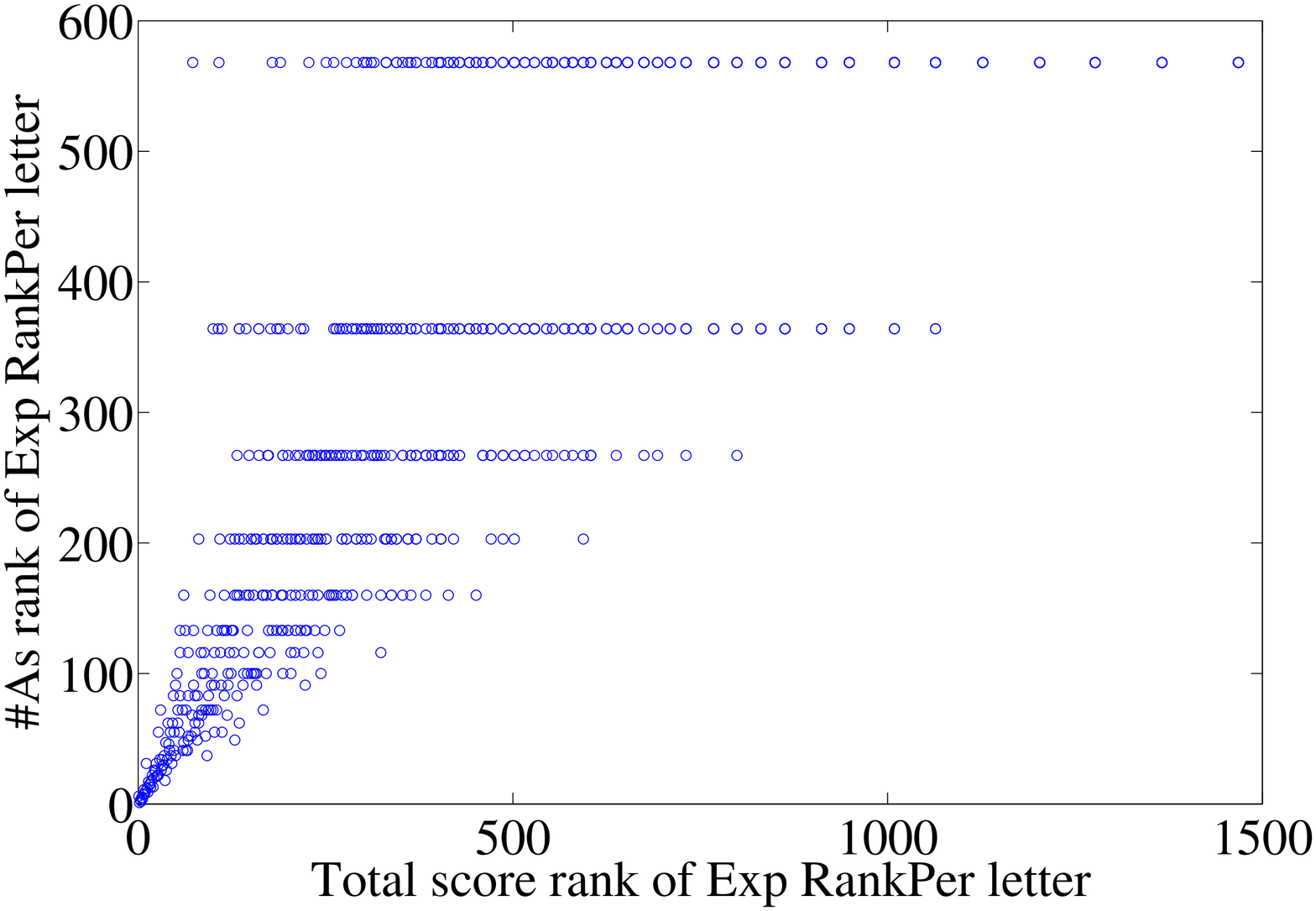}}
\caption{Comparison on institution ranking results between two
granularity methods: (counting number of As, y-axis) versus
(counting ``A''=1, ``B''=0.5, ``C''=0.25 for total score, x-axis)
for three metrics, Citation Count (CC), Influence (inf) and Exposure
(Exp) according to authors' rank percentile based letter
grades.}\label{Fig:RPtmVsRPal}
\end{figure*}

\begin{figure*}[hbt]
    \centering
    \subfigure[CC]{
    \label{Fig:ccRPtmVcctm}
    \includegraphics[width=1.4in]{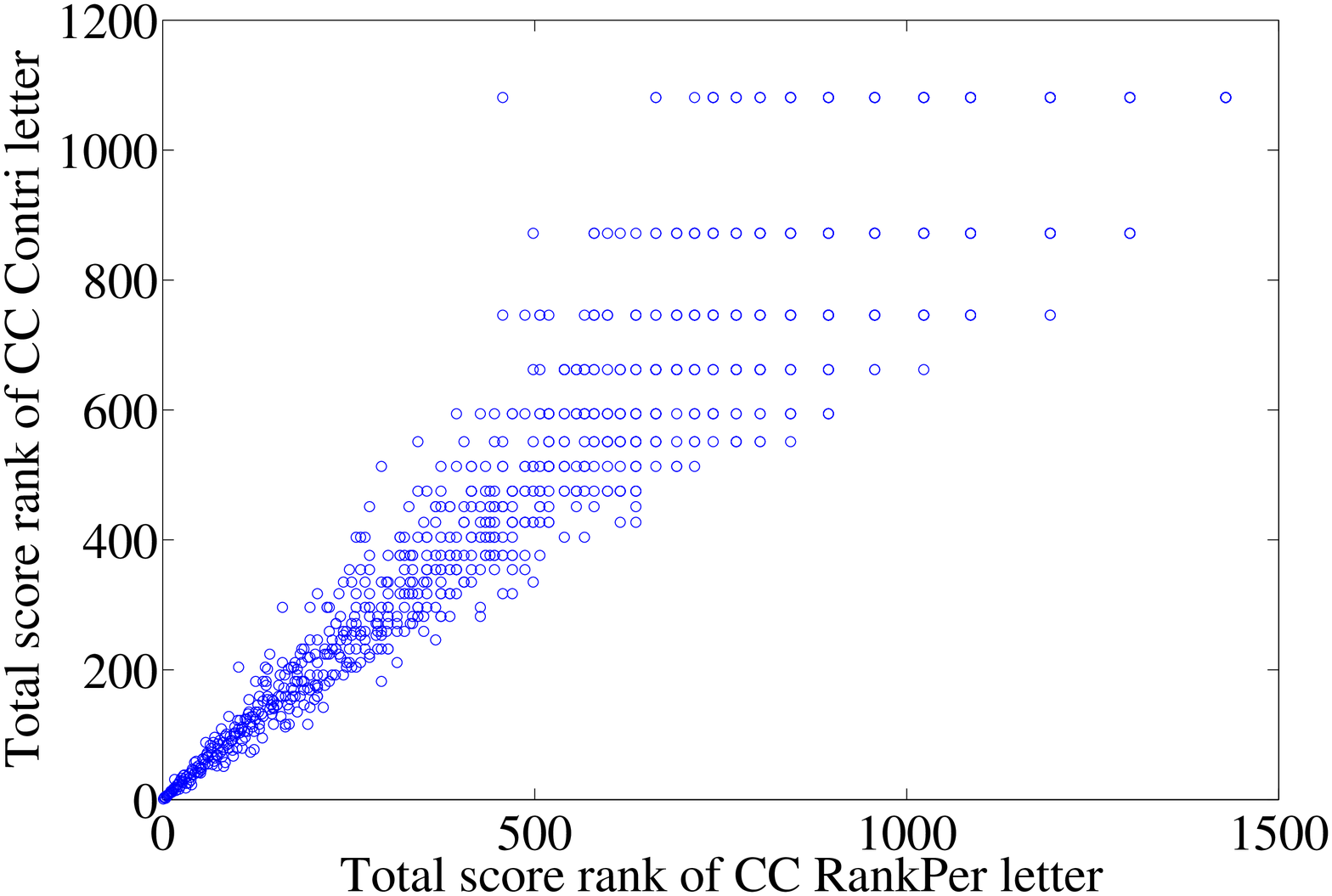}}
    \subfigure[Inf]{
    \label{Fig:infRPtmVinftm}
    \includegraphics[width=1.4in]{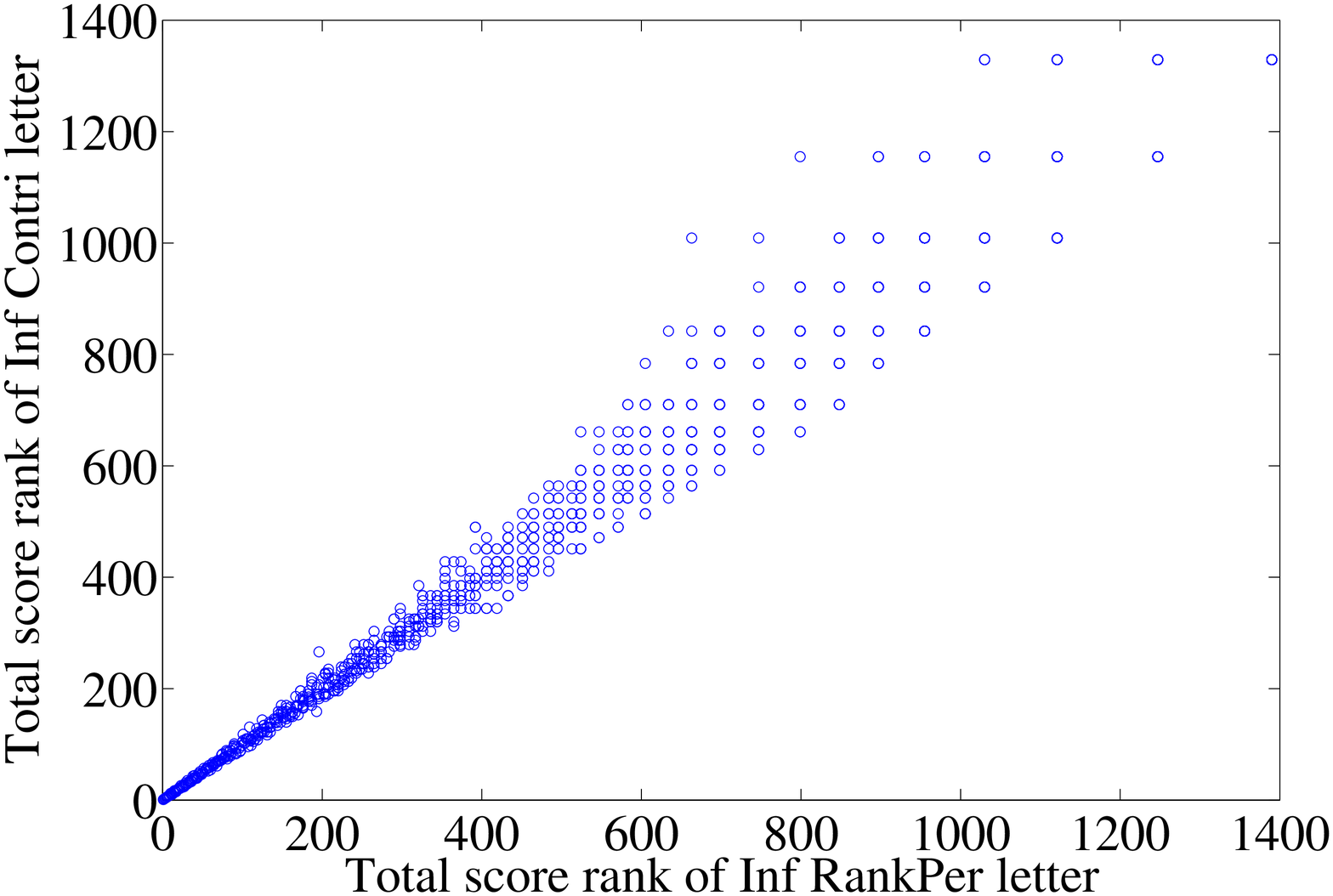}}
    \subfigure[Exp]{
    \label{Fig:avaRPtmVavatm}
    \includegraphics[width=1.4in]{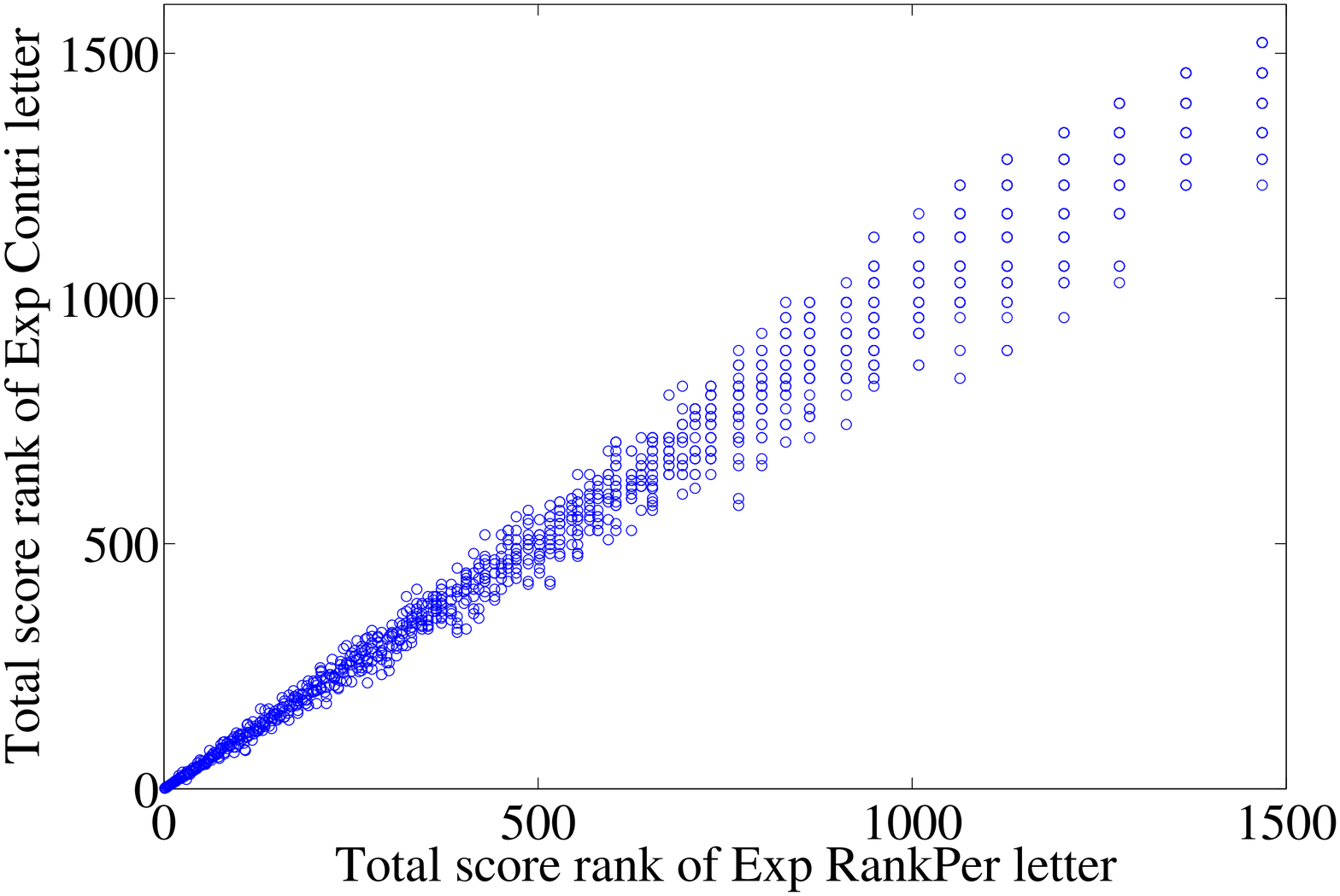}}
\caption{Comparison on institution ranking results between rank
percentile based (x-axis) versus contribution based (y-axis) letter
grading methods, using granularity method (2) for three metrics,
Citation Count (CC), Influence (inf) and Exposure (Exp).}
    \label{Fig:RPtmVstm}
\end{figure*}

\begin{figure*}[hbt]
    \centering
    \subfigure[Inf vs. CC]{
    \label{Fig:infRPtmVccRPtm}
    \includegraphics[width=2.1in]{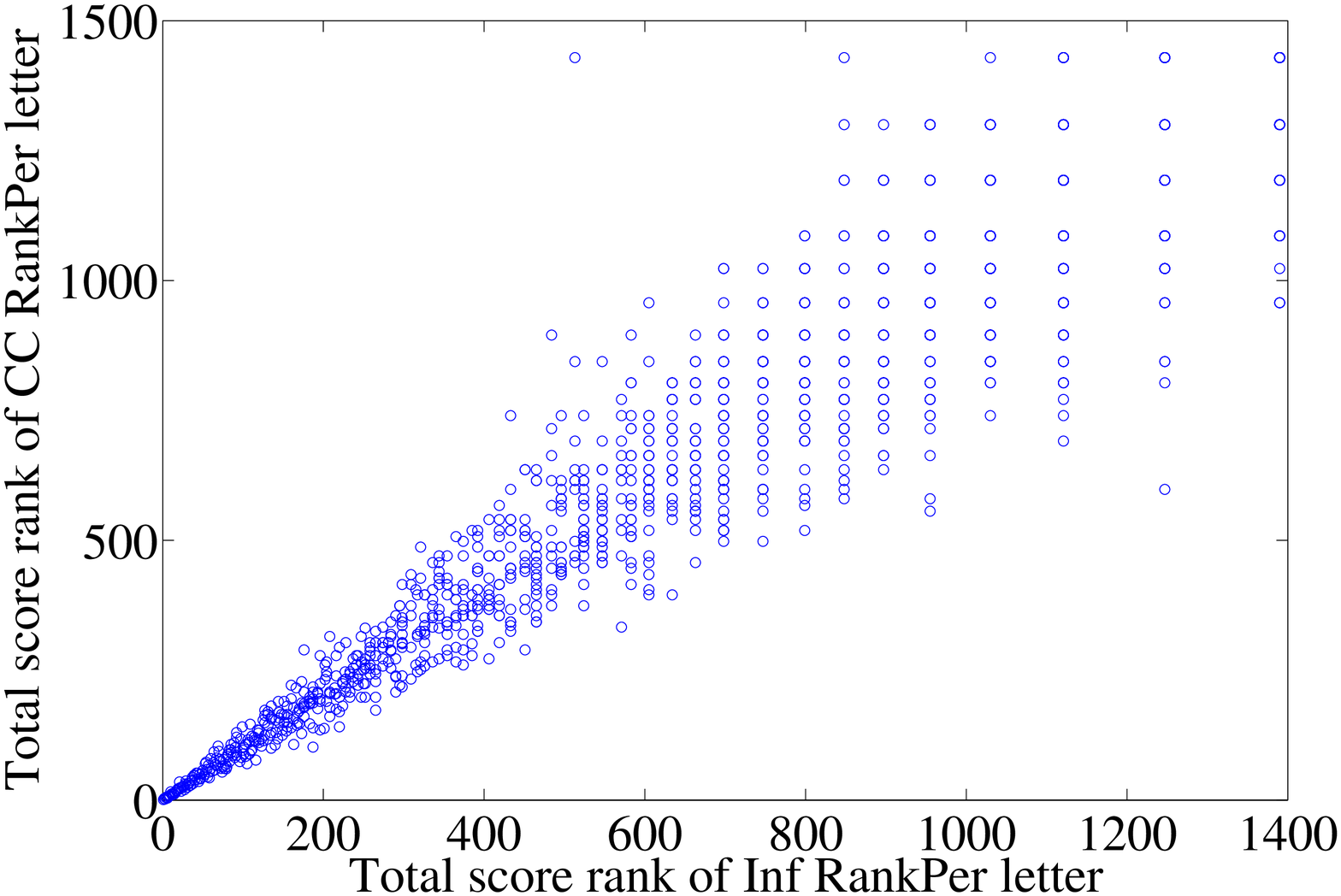}}
    \subfigure[Inf vs. Exp]{
    \label{Fig:infRPtmVavaRPtm}
    \includegraphics[width=2.1in]{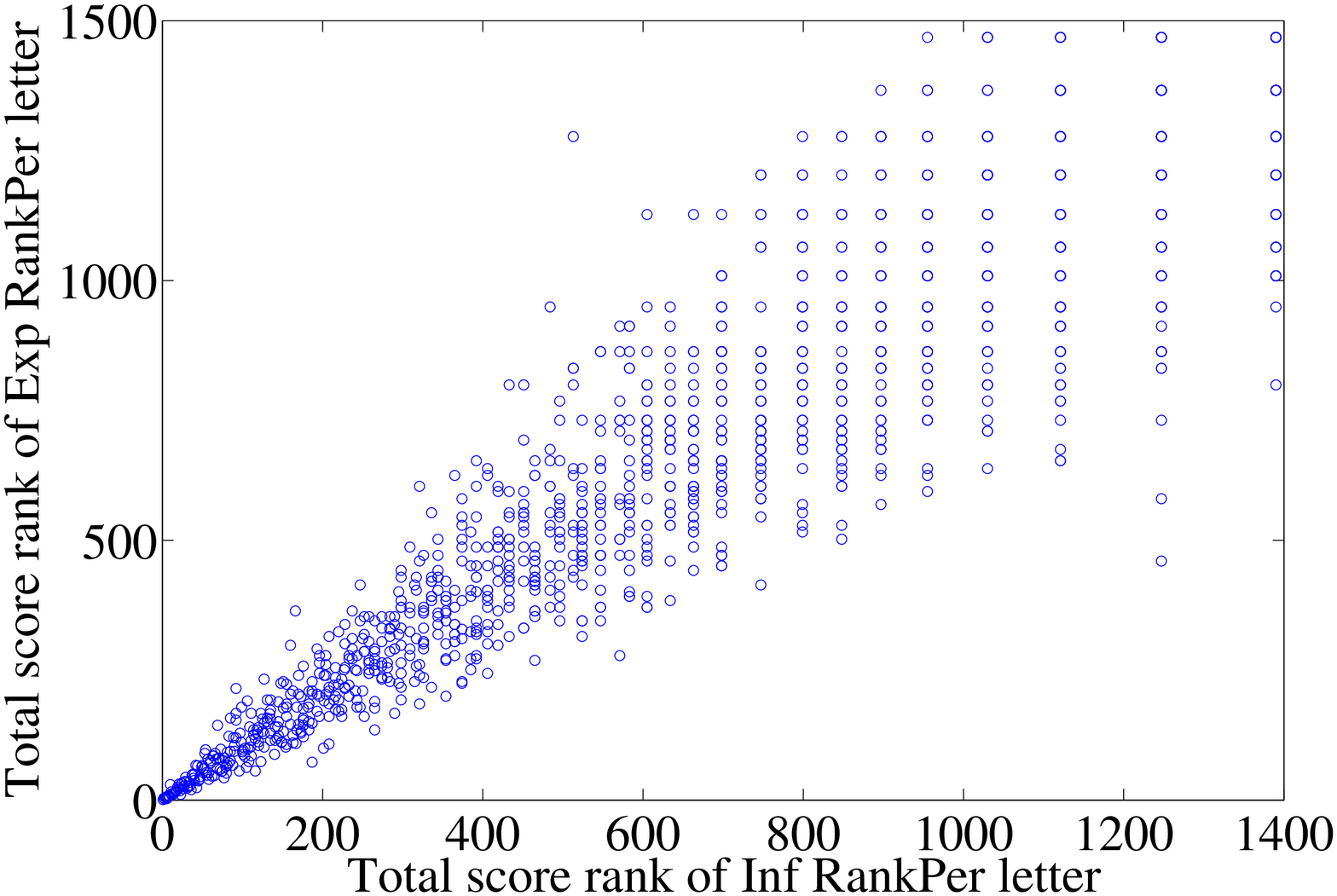}}
\caption{Comparison on institution ranking results among three
metrics, Citation Count (CC), Influence (inf) and Exposure (Exp)
according to rank percentile based letter grading results using
granularity method (2).}
    \label{Fig:RPtmVsRPtm}
\end{figure*}

Next we show three sets of similarity study between different
institution ranking results, mainly focused on three selected
metrics: Citation Count (CC), Influence (Inf) and Exposure (Exp). In
the first set, we compared the ranking results at two different
granularities, count number of ``A'' authors vs. compute total
score, based on the rank percentile based letter grades, as shown in
Fig.~\ref{Fig:infRPtmVinfRPal}. In the second set
(Fig.~\ref{Fig:RPtmVstm}), we investigated how the authors' letter
grading methods (rank percentile based vs. contribution based)
affect the total scores (granularity method (2)) as well as the
institution rankings. In the last set, we compared three metrics
(Inf vs. CC in Fig.~\ref{Fig:infRPtmVccRPtm}, Inf vs. Exp in
Fig.~\ref{Fig:infRPtmVavaRPtm}) while the rank percentile based
letter grading scheme and the granularity method (2) of computing
total score are used.

According to the above comparison results
(Figures~\ref{Fig:RPtmVsRPal},~\ref{Fig:RPtmVstm}
and~\ref{Fig:RPtmVsRPtm}), we made several observations:
\begin{enumerate}
\item[i.] As shown in Fig.~\ref{Fig:RPtmVsRPal}, for those highly
ranked institutions (e.g. above 100th), the ranking results of the
two granularities are very close. In addition, as mentioned before,
when the total scores (by counting ``A''=1, ``B''=0.5, ``C''=0.25)
were same, granularity method (1) can indicate the ratio of authors
earning letter ``A'' (e.g. Princeton University
in Table~\ref{Tab:AffRanking}).\\

On the other hand, counting the number of ``A'' authors only was
ineffective in distinguishing institutions ranked below 100th (note
the number of institutions with the same number of ``A''
authors located on horizontal lines).\\

\item[ii.] As shown in Fig.~\ref{Fig:RPtmVstm},
although the rank percentile based and contribution based letter
grading methods make pronounced differences on author rankings, they
produce very similar results on institution rankings.\\

\item[iii.] As shown in Fig.~\ref{Fig:RPtmVsRPtm}, institutions
ranked above 100th have similar ranking results for these three
metrics (CC, Inf and Exp); however, the points are spread out
largely for those ranked below 100th under different metrics. This
again validates the effectiveness of the definitions of the various
metrics with practical interpretations.
\end{enumerate}

Finally, we compare our institution ranking approach to the three
established ranking systems. We show the top 30 universities in
``Computer Science'' domain ranked by each of these systems,
together with the ranking results by ours (based on total score of
Influence metric):
\begin{enumerate}
\item[a)] US News Ranking - The
Best Graduate Schools in Computer Science ranked in
2010~\cite{USNewRank}. Results are shown in
Table~\ref{Tab:AffRankingUSN}.
\item[b)] The QS World University Rankings By Subject 2013 - Computer Science \& Information
Systems~\cite{QSCSRank}. Results are shown in
Table~\ref{Tab:AffRankingQS}.
\item[c)] The Academic Ranking of World Universities (ARWU by SJTU) 2012 in Computer
Science~\cite{ARWURank}. Results are shown in
Table~\ref{Tab:AffRankingSTJU}.
\end{enumerate}

As in Tables~\ref{Tab:AffRankingUSN}-\ref{Tab:AffRankingSTJU}, we
find that the calculation of the overall score is the key factor
leading to the deviation of the ranking results among different
systems. In particular, the US New ranking system applied a
subjective based approach~\cite{USNewRank} to calculate the total
scores for each university. The QS ranking system calculated the
overall score in the ``Computer Science \& Information Systems''
subject based on the four objective factors: ``Academic
Reputation'', ``Employer Reputation'', ``Citations per Paper'' and
``H-index Citations''~\cite{QSCSRank}. The ARWU ranking system, on
the other hand, consider the overall score in ``Computer Science''
domain as the weighted average of the five metrics: ``Alumni Turing
Awards ($10\%$)'', ``Staff Turing Award ($15\%$)'', ``Highly Cited
Researchers ($25\%$)'', ``Papers indexed in SCI ($25\%$)'' and
``Papers Published in Top Journals ($25\%$)''~\cite{ARWURank}.
Because of these factors (considering more reputation and recent
work), the results of the three ranking systems tend to be quite
volatile - the top universities change quite a bit from year to
year. In our case of using total score of Influence metric, we are
at least more stable and pure.

As Microsoft Libra also provides the institution ranking
services~\cite{libra}, we make another comparison and the results
are shown in Table~\ref{Tab:AffRankingLibra}. Since we are using the
same dataset for calculation, it is not surprising that the ranking
results are very similar.
%%%%%%%%%%%%%%%%%%%%%%%%%%%%%%%%%%%%%%%%
%%USNews
\newpage
\begin{table*}[!htb]
\centering \caption{Top 30 Universities of ``US News Ranking - The
Best Graduate Schools in Computer Science ranked in 2010'', compared
to ours (total score of Influence metric, \emph{Inf TS})} \label{Tab:AffRankingUSN}
\begin{tabular}{|c|c|c|c|c|c|c|}
%\hline
%
\hline University Name & Score & USNews &  Inf TS\\
\hline Carnegie Mellon University               & 5.0 & 1 & 2\\
\hline Massachusetts Institute of Technology    & 5.0 & 1 & 1\\
\hline Stanford University                      & 5.0 & 1 & 3\\
\hline University of California Berkeley        & 5.0 & 1 & 4\\
\hline Cornell University                       & 4.6 & 5 & 14\\
\hline University of Illinois Urbana Champaign  & 4.6 & 5 & 5\\
\hline University of Washington                 & 4.5 & 7 & 9\\
\hline Princeton University                     & 4.4 & 8 & 28\\
\hline University of Texas Austin               & 4.4 & 8 & 11\\
\hline Georgia Institute of Technology          & 4.3 & 10 & 6\\
\hline California Institute of Technology       & 4.2 & 11& 33\\
\hline University of Wisconsin Madison          & 4.2 & 11 & 17\\
\hline University of Michigan                   & 4.1 & 13 & 13\\
\hline University of California Los Angeles     & 4.0 & 14 & 11\\
\hline University of California San Diego       & 4.0 & 14  & 8\\
\hline University of Maryland                   & 4.0 & 14 & 9\\
\hline Columbia University                      & 3.9 & 17 & 16\\
\hline Harvard University                       & 3.9 & 17 & 23\\
\hline University of Pennsylvania               & 3.9 & 17 & 20\\
\hline Brown University                         & 3.7 & 20& 42\\
\hline Purdue University                        & 3.7 & 20 & 25\\
\hline Rice University                          & 3.7 & 20 & 47\\
\hline University of Massachusetts              & 3.7 & 20 & 25\\
\hline University of North Carolina-Chapel Hill & 3.7 & 20 & 42\\
\hline University of Southern California        & 3.7 & 20 & 6\\
\hline Yale University                          & 3.7 & 20 & 53\\
\hline Duke University                          & 3.6 & 27 & 59\\
\hline Johns Hopkins University                 & 3.4 & 28 & 44\\
\hline New York University                      & 3.4 & 28 & 33\\
\hline Ohio State University                    & 3.4 & 28 & 40\\
\hline Pennsylvania State University            & 3.4 & 28 & 46\\
\hline Rutgers, The State University of New Jersey  & 3.4 & 28 & 21\\
\hline University of California Irvine          & 3.4 & 28 & 24\\
\hline University of Virginia                   & 3.4 & 28 & 68\\
\hline
\end{tabular}
\end{table*}

%%%%%%%%%%%%%%%%%%%%%%%%%%
%%QS
\newpage
\begin{table*}[!htb]
\centering \caption{Top 30 Universities of ``The QS World University
Rankings By Subject 2013 - Computer Science \& Information
Systems'', compared to ours (total score of Influence metric, \emph{Inf TS})}
\label{Tab:AffRankingQS}
\begin{tabular}{|c|c|c|c|c|c|c|}
%\hline
%
\hline University Name                          & Score & QS &  Inf TS\\
\hline Massachusetts Institute of Technology    & 96.7  & 1 & 1\\
\hline Stanford University                      & 92.1  & 2 & 3\\
\hline University of Oxford                     & 92.0  & 3 & 21\\
\hline Carnegie Mellon University               & 90.5  & 4 & 2\\
\hline University of Cambridge                  & 89.8 & 5 & 15\\
\hline Harvard University                       & 88.4 & 6 & 23\\
\hline University of California Berkeley        & 88.0 & 7 & 4\\
\hline National University of Singapore         & 87.2 & 8 & 57\\
\hline Swiss Federal Institute of Technology Zurich  & 87.1 & 9 & 22\\
\hline University of Hong Kong                 & 84.0 & 10 & 165\\
\hline Princeton University                     & 83.7 & 11 & 28\\
\hline The Hong Kong University of Science \& Technology & 83.6 & 12 & 113\\
\hline The University of Melbourne              & 83.4 & 13 & 82\\
\hline University of California Los Angeles     & 82.1 & 14 & 11\\
\hline University of Edinburgh                  & 81.5 & 15 & 30\\
\hline University of Toronto                    & 81.0 & 16 & 18\\
\hline \'{E}cole Polytechnique F\'{e}d\'{e}rale de Lausanne & 80.2 & 17  & 36\\
\hline Imperial College London                  & 79.7 & 18 & 35\\
\hline The Chinese University of Hong Kong      & 79.5 & 19 & 94\\
\hline The University of Tokyo                  & 79.4 & 20 & 50\\
\hline Australian National University           & 78.9 & 21& 107\\
\hline Nanyang Technological University         & 78.5 & 22 & 91\\
\hline University College London                & 78.0 & 23 & 47\\
\hline The University of Sydney                 & 77.9 & 24 & 146\\
\hline The University of Queensland             & 77.8 & 25 & 107\\
\hline Cornell University                       & 77.6 & 26 & 14\\
\hline Tsinghua University                      & 77.5 & 27 & 107\\
\hline University of Waterloo                   & 77.5 & 27 & 32\\
\hline The University of New South Wales        & 77.3 & 29 & 102\\
\hline The University of Manchester             & 77.1 & 30 & 45\\
\hline
\end{tabular}
\end{table*}

%%%%%%%%%%%%%%%%%%%%%%%%%%%%%%%%%%%%%%
%%SJTU-ARWU
\newpage
\begin{table*}[!htb]
\centering \caption{Top 30 Universities of ``The Academic Ranking of
World Universities (ARWU by SJTU) 2012 in Computer Science, compared
to ours (total score of Influence metric, \emph{Inf TS})}
\label{Tab:AffRankingSTJU}
\begin{tabular}{|c|c|c|c|c|c|c|}
%\hline
%
\hline University Name                          & Score & SJTU & Inf TS\\
\hline Stanford University                      & 100 & 1 & 3\\
\hline Massachusetts Institute of Technology    & 93.8 & 2 & 1\\
\hline University of California Berkeley        & 85.3 & 3 & 4\\
\hline Princeton University                     & 78.7 & 4 & 28\\
\hline Harvard University                       & 77.7 & 5 & 23\\
\hline Carnegie Mellon University               & 71.8 & 6 & 2\\
\hline Cornell University                       & 71.2 & 7 & 14\\
\hline University of California Los Angeles     & 69.2 & 8 & 11\\
\hline University of Texas Austin               & 68.3 & 9 & 11\\
\hline University of Toronto                    & 63.6 & 10 & 18\\
\hline California Institute of Technology       & 63.5 & 11 & 33\\
\hline Weizmann Institute of Science            & 63.3 & 12 & 89\\
\hline University of Southern California        & 63.0 & 13 & 6\\
\hline University of California San Diego       & 61.8 & 14  & 8\\
\hline University of Illinois Urbana Champaign  & 61.7 & 15 & 5\\
\hline University of Maryland                   & 60.1 & 16 & 9\\
\hline University of Michigan                   & 58.9 & 17 & 13\\
\hline Technion-Israel Institute of Technology  & 57.8 & 18 & 29\\
\hline University of Oxford                     & 56.7 & 19& 31\\
\hline Purdue University                        & 54.5 & 20 & 25\\
\hline University of Washington                 & 54.2 & 21 & 9\\
\hline Columbia University                      & 53.8 & 22 & 16\\
\hline Rutgers, The State University of New Jersey  & 53.5 & 23 & 21\\
\hline Georgia Institute of Technology          & 53.0 & 24 & 6\\
\hline Swiss Federal Institute of Technology Zurich & 52.7 & 25& 22\\
\hline The Hong Kong University of Science \& Technology & 52.6 & 26 & 113\\
\hline The Hebrew University of Jerusalem       & 52.5 & 27 & 77\\
\hline Yale University                          & 51.4 & 28& 53\\
\hline Tel Aviv University                      & 50.9 & 29& 36\\
\hline The Chinese University of Hong Kong      & 50.7 & 30& 94\\
\hline
\end{tabular}
\end{table*}

%%%%%%%%%%%%%%%%%%%%%%%%%%%%%%%%%%%%%%
%%Libra
\newpage
\begin{table*}[!htb]
\centering \caption{Top 30 Universities ranked by Libra in
``Computer Science'' domain, compared to ours (total score of
Influence metric, \emph{Inf TS})} \label{Tab:AffRankingLibra}
\begin{tabular}{|c|c|c|c|c|c|c|}
%\hline
%
\hline University Name & Field Rate & Libra &  Inf TS\\
\hline Stanford University                      & 418 & 1 & 3\\
\hline Massachusetts Institute of Technology    & 408 & 2 & 1\\
\hline University of California Berkeley        & 404 & 3 & 4\\
\hline Carnegie Mellon University               & 325 & 4 & 2\\
\hline University of Illinois Urbana Champaign  & 268 & 5 & 5\\
\hline Cornell University                       & 260 & 6 & 14\\
\hline University of Southern California        & 256 & 7 & 6\\
\hline University of Washington                 & 256 & 7 & 9\\
\hline University of California San Diego       & 253 & 9  & 8\\
\hline Princeton University                     & 252 & 10 & 28\\
\hline University of Texas Austin               & 248 & 11 & 11\\
\hline University of California Los Angeles     & 243 & 12 & 11\\
\hline University of Maryland                   & 238 & 13 & 9\\
\hline Georgia Institute of Technology          & 229 & 14 & 6\\
\hline University of Michigan                   & 224 & 15 & 13\\
\hline University of Toronto                    & 222 & 16 & 18\\
\hline University of Cambridge                  & 214 & 17 & 15\\
\hline Harvard University                       & 214 & 17 & 23\\
\hline University of Wisconsin Madison          & 209 & 19 & 17\\
\hline Columbia University                      & 202 & 20 & 16\\
\hline University of Pennsylvania               & 201 & 21 & 20\\
\hline University of California Irvine          & 199 & 22 & 24\\
\hline Rutgers, The State University of New Jersey  & 197 & 23& 21\\
\hline University of Oxford                     & 197 & 23& 31\\
\hline University of Minnesota                  & 195 & 25& 25\\
\hline Swiss Federal Institute of Technology Zurich & 190 & 26& 22\\
\hline The French National Institute for        & 189 & 27& 19\\
Research in Computer science and Control        & &   &   \\
\hline California Institute of Technology  & 189 & 27& 33\\
\hline Brown University                         & 189 & 27& 42\\
\hline University of Massachusetts              & 189& 27& 25\\
\hline
\end{tabular}
\end{table*}

%
%\section{Related works}
%\input{related}
%%%Sextion related works
\section{Related Works}
The study of academic publication statistics is by no means a new
topic. Previous attention focused mostly in different areas of
science, especially physics. The most influential work was published
in 1965 by Derek \cite{DJSP_science}, in which he considered papers
and citations as a network and noticed the citation distribution
(degree distribution) followed the power law. A few years later, he
tried to explain this phenomenon using a simple model called the
\emph{cumulative advantage} process
Derek~\cite{price1976,RKM_science}. The skewness of the citation
count distribution has since been validated by other studies on
large scale datasets~\cite{Seglen1992,SR_EurPhys}. In subsequent
literature, later on, the model became better known as
\emph{preferential attachment} by \cite{BA_Science1999} (i.e. a
paper is more likely to cite another paper with more existing
citations) and with good empirical evidence~\cite{JNB_EurLett2003}.

To determine the quality or \emph{impact} of a paper by its citation
count, while considered reasonable by many, has met with strong
criticisms~\cite{WBHF_2003}. Instead of using citation count,
it has been proposed that a ranking factor, calculated using the
eigenvector-based methods such as PageRank~\cite{pagerank} or
HITS~\cite{hits}, be adopted. Subsequently, a number of proposals of
different variations to measure paper importance appeared, including
eigenvector-based~\cite{Sun_Giles_2007,eigenfactor} or network traffic-like
schemes~\cite{WXYM_2007,LiYLHD11}. Since it takes time for a paper to
accumulate its share of citations, it is
common practice to use the venue (journal) the paper is published in
to predict the potential impact/importance of a paper. Thus, Journal
Impact Factor (JIF~\cite{jif}) becomes an important indicator used
in practice.

The use of citation count has become more popular due to Google
Scholar. More recently, some new indices, such as
h-index~\cite{hindexNature,hindex} and g-index~\cite{gindex} have
been proposed to combine the use of citation count and paper count
to measure the achievements of an author. Some recent studies have also proposed to apply PageRank-type
iterative algorithms to evaluate authors' contribution and impact,
notably a scheme called SARA (Scientific Author Ranking Algorithm)
to compute authors contributions~\cite{sara}; and a model to rank
both papers and authors~\cite{leeGiles_coranking}.

Besides the paper citations \emph{earned} by authors, authors can
also be ranked based on their connections and popularity as a
co-author. This way of evaluating authors is used in a series of
studies by Newman \emph{et al} on author collaboration
networks~\cite{MEJNewman_PhysRevE2001,MEJNewman_PNAS2001,MEJNewman_PNAS2004,MEJNewman_Springer2004}.
This approach and viewpoint is similar to that used in the study of
social networks~\cite{Kleinberg_Book_2010}. A number of recent
papers studied social influence and their correlation to user
actions \cite{BKA_EC09, AKM_KDD08,
CCHKS_KDD08,budalakoti2012bimodal}.

Finally, the publication database plays a critical role in such
bibliometrics and social network studies. The well-known databases
are: Google Scholar, Scopus, ISI, CiteSeer~\cite{citeseer},
Microsoft Libra~\cite{libra}, DBLP~\cite{dblp}, IEEE, ACM. These
databases, however, tend to contain different
papersets~\cite{CF_CCR_2010}. For example, CiteSeer, DBLP, ACM focus
mostly on computer science and related literature, but each has its
own rules of which conferences/papers to include or not. Not all
these databases have citation information (e.g. DBLP does not).
%

%
%\input{discussions}
%%section:discussions
\section{Discussions}

\subsection{The name disambiguation problem}
Name ambiguity is a big problem with online systems dealing with people names
without explicit registration, especially true for bibliometric systems since
the publication records come from many years of accumulation and from many
different publishers. It is a hard problem, the full solution of which is
beyond the scope of this paper. Below, we discuss some of the steps that
have been taken and our plans for dealing with this problem in the future.

\begin{figure}[h]
    \centering
    \includegraphics[width=2.3in]{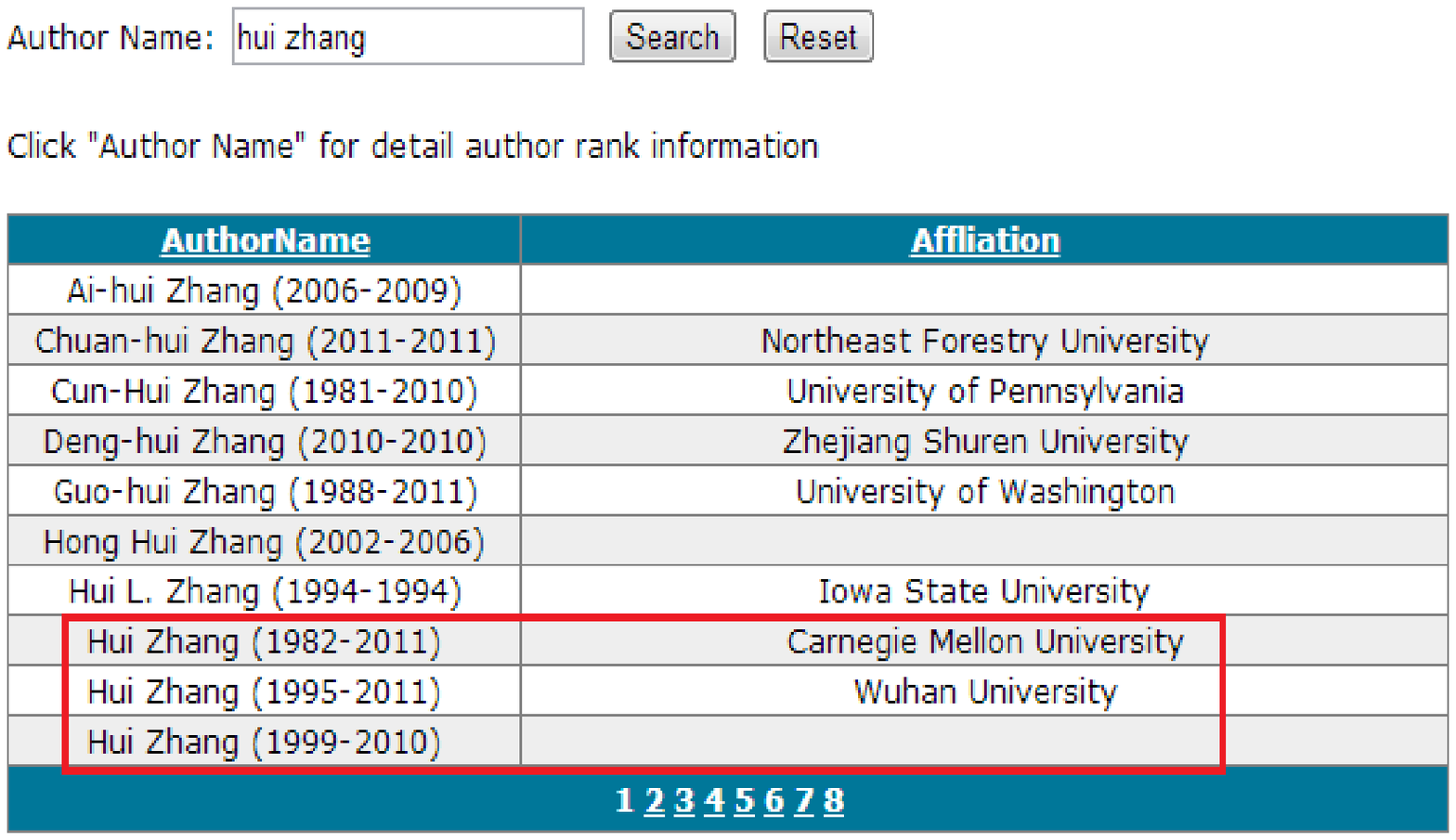}
    %\label{namedis1}
%\end{figure}
%\begin{figure}[h]
    %\centering
    \includegraphics[width=2.3in]{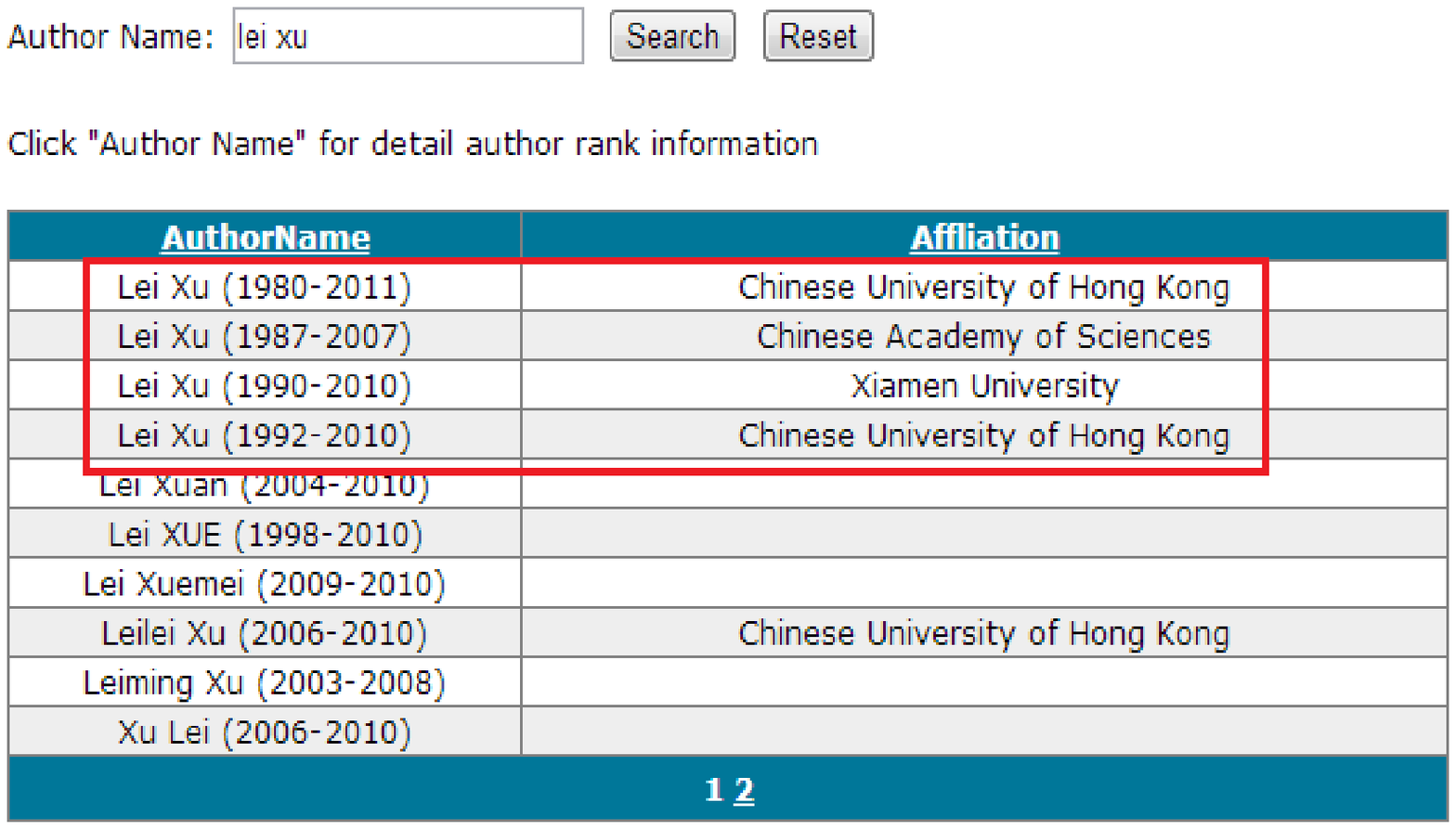}
%    \caption{Comparison between Influence and the year of first publication.}
    \caption{Examples of name disambiguation results.}
    \label{namedis1}
\end{figure}

Our current implementation of the \emph{Academic Influence Ranking} system makes
full use of the objectized data from Microsoft~\cite{libra}. Each author is an object
with its own ID. Microsoft Libra has already applied some name disambiguation algorithm
to clean its raw data. We show two examples to illustrate this in Fig.~\ref{namedis1}.

As shown by the examples, multiple authors with the same name but different affiliations are
included in Libra's dataset, and we access the authors by their IDs.

From examining specific cases, we know that there still exist many author names (and their IDs)
that are shared by many different real-world persons. MS Libra is also aware of this problem,
evidenced by the fact that they submitted this problem as a challenge for the~\cite{kddcup2013}.
We expect MS Libra will apply the algorithms proposed by the winning team of this competition in
the near future. Since we plan to continue to update our system by sourcing data from MS Libra,
we need to be careful in doing our own name disambiguation so that we can continue to leverage
of the MS Libra data.

On the other hand, we are also using our tool and dataset for various statistical analysis,
and model validation. For such purposes, it is sometimes adequate to disambiguate only the authors
with significant publications. For this we can apply some semi-automatic and semi-manual methods.
For example, we can automatically identify the author names worthy of disambiguation, and do the
disambiguation semi-manually. Here are some semi-manual methods we are trying:
\begin{enumerate}
\item[1)] We have developed a crawler-parser to extract online information (e.g. author's homepage)
for given author names, and use that information to disambiguate authors with the same name.

\item[2)] We have also found certain online services with author registration, that can potentially
help us disambiguate authors manually.\\
\end{enumerate}

This allows us to be more confident with our statistical inferences.

In the long run, we believe the ultimate solution requires us to based everything on an (single)
author registration system, so that all authors are guaranteed unique. This is clearly not a technical issue any more.

\subsection{User feedbacks}
We have demonstrated our system to many colleagues and friends, including some experts from the industry (Elsevier).
Overall, we received very positive feedback. Here are some things people liked a lot:
\begin{enumerate}
\item[1)] By checking out the scores for authors familiar, the reviewers told us that the use of influence and connections seem to sort out the stronger researchers from those socially active researchers.

\item[2)] By checking the university ranking for domains familiar to them, the reviewers told us that the ranking is quite accurate, and the top universities are exactly the ones with strong groups in that domain.

\item[3)] Many told us that our website can be very useful for: (i) students searching for finding supervisors and graduate programs to apply; (ii) TPC chairs or journal editors finding people to review papers; (iii) hiring search; (iv) occasionally checking out someone to get their relative position roughly.
\end{enumerate}

We also received many good suggestions that we will follow up in our future works. Here are some example ones:
\begin{enumerate}
\item[1)] It would be good to do controlled survey of (systematically selected) people in the different field, to see their opinions.

\item[2)] It would be good to introduce the concept of peer group for each person, and do comparison in that context. For example, a person's peer group should include people of similar years of research experience.

\item[3)] It would be important to develop the user feedback component into the current website.
\end{enumerate}

%\section{Conclusion}
%\input{conclusions}
%%section:conclusion
\section{Conclusion}
In this paper, we present the design and experimental study of an
Academic Social Network website (\url{http://pubstat.org}) that we
have built. It consists of several different non-conventional,
social-network-like metrics we can use to rank authors and compare
authors.
%% newly added start
In addition, it also provides author-based institution rankings by
utilizing the author-institution relationship information.
%% newly added end
%It has been demonstrated to numerous colleagues and collaborators,
%many of whom found it very useful.
It has been demonstrated to many colleagues and friends, including some experts from industry (Elsevier).
Overall, we received very positive feedback and many good suggestions that we will follow up in our future works.

Although we have had a working system for some time now, there are
still many challenges to making it widely used. The publications
database we have is not as complete as we would like; and we want to
work out a way to continuously update it. The data is also far from
\emph{clean}. We are starting new projects to apply machine learning
techniques to \emph{clean} the data (some preliminary results in
estimating missing years on papers have been submitted for
publication).

We continue to discover new query types that users are interested
in, and even new metrics. If the reviewers of this paper are
interested in examining our website, we would be glad to open it for
inspection in some fashion (\url{http://pubstat.org}).

\bibliographystyle{spbasic}      % basic style, author-year citations
\bibliography{AcademicAuthroRanking}
\appendix
%\input{appendix}
%%appendix
\section{The PageRank Algorithm}\label{Sec:AppendixA}

Given a graph $G=(V,E)$, the PageRank Algorithm can be considered as
a random walk starting from any node along the edges. After an
infinite number of steps, the probability that a node is visited is
the PageRank value of that node.

More formally, the probability distribution of visiting each node
can be derived by solving a Markov Chain. The transition matrix
$C$'s entries $c_{ij}$ ($i,j=1,2,\dots, n$) represent the transition
probability that the random walk will visit node $j$ next given that
it is currently at node $i$. Thus, $c_{ij}$ can be expressed as
\begin{equation}
\label{eq:pr} c_{ij} = Prob(j|i) = \frac{e_{ij}}{\sum_k e_{ik}}
\end{equation}
where $e_{ij}$ is from the adjacency matrix for the graph $G$. If
$G$ is the citation graph, for example, then $e_{ij}=1$ if paper $i$
cites paper $j$; else $e_{ij}=0$.

In general, $C$ is a {\em substochastic} matrix with rows summing to
either 0 (dangling nodes~\cite{pagerank}, for example, representing
papers with citing no other papers) or 1 (normal nodes, or papers).
For each dangling node, the corresponding row is replaced by
$\frac{1}{n}\mathbf{e}$, so that $C$ becomes a {\em stochastic}
matrix.

In order to ensure the Markov Chain $C$ is irreducible, hence a
solution is guaranteed to exist, $C$ is further transformed as
follows:
\begin{equation}
\label{eq:prp} \widetilde{C} = \alpha C +
(1-\alpha)\mathbf{e}\mathbf{v}^T,\;\;\alpha\in(0,1).
\end{equation}
Here, $\mathbf{e}$ is a special column vector with all 1s, and of
dimension $n$.

In Eq.~(\ref{eq:prp}), $\mathbf{v}\in\mathcal{R}^{n}$ is a
probability vector (i.e. its values are between $0$ and $1$, and sum
to $1$). It is referred to as the {\em teleportation vector}, which
can be used to configure some bias into the random walk. For our
purposes, we let $\mathbf{v} = 1/n\mathbf{e}$ as the default
setting.

Now, according to the Perron-Frobenius Theorem~\cite{Meyer_PageRank,
Meyer_SIAM2000}, matrix $\widetilde{C}$ is {\em stochastic,
irreducible} and {\em aperiodic}, and the equation
\begin{eqnarray}
\label{eq:pi} \mathbf{\pi}^T = \alpha\mathbf{\pi}^T C +
(1-\alpha)\frac{1}{n}\mathbf{e}^T,\;\;\alpha\in(0,1)
\end{eqnarray}
which can be solved by iteration methods in practice.

%\section{Definition of Metrics in Matrix Form}\label{Sec:AppendixB}
%

%\input{appendixB}
%%appendix
\section{Definition of Metrics in Matrix Form}\label{Sec:AppendixB}
We list the matrix form for the 5 metrics discussed in the previous
sections in the following table:
\begin{table}[thb]
\centering
\caption{Notations and derivations of the ranking metrics.}
\begin{tabular}{|r|l|}
\hline
Notations & Description\\
\hline
 $n_P$  & total number of papers\\
\hline
 $n_A$  & total number of authors\\
\hline
 $n_V$  & total number of venues\\
\hline
 $(X)^*$ & row normalization operation on any $X$, i.e.,\\
         & $(X)_{ij}^* = \frac{X_{ij}}{\sum_k X_{ik}}$, for non-zero rows\\
\hline
 $R$    & $n_P \times n_P$ paper-citation adjacent matrix,\\
        & $R_{ij} = 1$, if paper $i$ has cited paper $j$\\
        & 0, otherwise.\\
\hline
 $A$    & $n_P \times n_A$ paper-author adjacent matrix,\\
        & $A_{ij} = 1$, if paper $i$ is written by author $j$\\
        & 0, otherwise.\\
\hline
 $V$    & $n_P \times n_V$ paper-venue adjacent matrix,\\
        & $V_{ij} = 1$, if paper $i$ has published in venue $j$\\
        & 0, otherwise.\\
\hline
 $H$    & $n_A \times n_A$ author influencing matrix,\\
        & $H = (A^T)^*(R)^*(A)^*$\\
\hline
 $Y$    & $n_V \times n_V$ venue influencing matrix,\\
        & $Y = (V^T)^*(R)^*(V)^*$\\
\hline
 $F$    & $n_A \times n_A$ author following indicating matrix,\\
        & $F_{ij} = 1$ if author $i$ has cited author $j$'s paper\\
        & at least once, else 0.\\
\hline
 $N$    & $n_A \times n_A$ author collaboration matrix,\\
        & $N = A^TA$\\
\hline
 $T_{VA}$   & $n_V \times n_A$ matrix, $T_{VA} = (V^T)^*(A)^*$\\
\hline
 $T_{AV}$   & $n_A \times n_V$ matrix, $T_{AV} = (A^T)^*(V)^*$\\
\hline
 $P$        & $(n_A + n_V) \times (n_A + n_V)$ matrix, \\
            & $P =  \left(
                        \begin{array} {c c}
                            \alpha (H)^* & (1 - \alpha)T_{AV} \\
                            (1 - \alpha)T_{VA} & \alpha (Y)^*
                        \end{array}
                    \right)$\\
\hline \hline
Metrics & Description\\
\hline
 CV         & apply PageRank on $(R)^*$ to get papers CV,\\
            & assign papers CV equally to authors,\\
            & through, $\pi^T (A)^*$\\
\hline
 Influence  & apply PageRank on $(H)^*$\\
\hline
 Follower   & apply PageRank on $(F)^*$\\
\hline
 Connection & apply PageRank on $(N)^*$\\
\hline
 Exposure  & apply PageRank on $P$, exposure of \\
           & both authors and venues are obtained\\
\hline

\end{tabular}
\label{Tab:AppNotation}
\end{table}

%\balancecolumns
\end{document}